# Layer engineered interlayer excitons

Qinghai Tan[1], Abdullah Rasmita[1], Si Li[2,3], Sheng Liu[1], Zumeng Huang[1], Qihua Xiong[4], Shengyuan A. Yang[2]*, K. S. Novoselov[5]*, Wei-bo Gao[1,6]*

[1]*Division of Physics and Applied Physics, School of Physical and Mathematical Sciences, Nanyang Technological University, 637371, Singapore*
[2]*Research Laboratory for Quantum Materials in Singapore University of Technology and Design Singapore 487372, Singapore*
[3]*School of Physics, Northwest University, Xi' an 710069, China*
[4]*State Key Laboratory of Low-Dimensional Quantum Physics and Department of Physics, Tsinghua University, Beijing, China*
[5]*Department of Materials Science & Engineering National University of Singapore, 9 Engineering Drive 1, 117575, Singapore*
[6]*The Photonics Institute and Centre for Disruptive Photonic Technologies, Nanyang Technological University, 637371, Singapore*

To whom correspondence should be addressed: shengyuan_yang@sutd.edu.sg，kostya@nus.edu.sg，wbgao@ntu.edu.sg

## Abstract

**Photoluminescence (PL) from excitons serves as a powerful tool to characterize optoelectronic property and band structure of semiconductors, especially for atomically thin 2D transition metal chalcogenide (TMD) materials. However, PL quenches quickly when the thickness of TMD material increases from monolayer to few-layers, due to the change from direct to indirect band transition. Here we show that PL can be recovered by engineering multilayer heterostructures, with the band transition reserved to be direct type. We report emission from layer engineered interlayer excitons from these multilayer heterostructures. Moreover, as desired for valleytronic devices, the lifetime, valley polarization, and the valley lifetime of the generated interlayer excitons can all be significantly improved as compared with that in the monolayer-monolayer heterostructure. Our results pave the way for controlling the properties of interlayer excitons by layer engineering.**

## Main

Layered transition metal dichalcogenides (TMDs) $MX_2$ (M=W, Mo, and X=S, Se) provide a platform for the fundamental research and optoelectronics devices using atomically thin material[1]. For the optoelectronic applications, to have efficient light emission, a semiconductor with direct bandgap nature is required[2-4]. However, bilayer or multilayer $MX_2$ has indirect bandgap, quenching its light emission to large extend[2,5]. In this work, we show that Van der Waals (vdW) heterostructures[6] formed by stacking of two different $MX_2$ can convert indirect bandgap material to direct type and therefore recover its photoluminescence (PL) emission.

In particular, the light emission of interlayer excitons (IXs)[7-9] in our monolayer (1L) $WSe_2$/few layers (mL) $MoS_2$ heterostructures shows fascinating valley properties[7-9], where the lifetimes, valley polarization and the valley relaxation lifetimes improves remarkably. Valley, as defined as local band extreme in momentum space, represents one potential degree of freedom to carry and process information[10-12] in valley spintronics. Therein, a long valley lifetime and high valley degree of polarization are desired. However, the picosecond timescale depolarization process of the intralayer excitons in the monolayer TMDs dramatically limits its application[13,14]. For the IXs in the bilayer heterostructures, the electrons and holes are separated in different layers, thus reducing the electron-hole exchange interaction. This spatially indirect nature of the IXs results in their high valley polarization and long lifetime[7,9,15-17], which brings promising opportunities for high-temperature exciton condensation[17,18] and valley physics[7,9,15-17,19-22]. Different knobs have been shown to be effective ways to tune the properties of IXs, for instance, the twist angles[19,23-27]. The IXs formed in a bilayer heterostructure with small twist angles leads to the Moiré physics[19,25-27]. Here we show that, in addition to controlling the stacking angle in a bilayer heterostructure system, varying the number of layers provides another natural degree of freedom to modulate the vdW heterostructure properties. Two types of heterostructure are analyzed with different layer dependent behaviours: few layers (mL) $WSe_2$/monolayer (1L) $MoS_2$ and monolayer (1L) $WSe_2$/few layers (mL) $MoS_2$. For 1L-$WSe_2$/mL-$MoS_2$, the increase of layer numbers reduces the electron and hole cloud overlap, further increasing the valley lifetime of excitons. While the valley polarization almost vanishes for mL-$WSe_2$/1L-$MoS_2$. In addition, we found that, for 1L-$WSe_2$/mL-$MoS_2$, a tiny magnetic field can suppress the valley relaxation and increase the valley lifetime from tens of ns by two orders to several microsecond. The suppression of valley relaxation has been directly observed from the time resolved PL emission due to the slow valley relaxation in these multilayer vdW heterostructures. In the end, a more robust valley polarization can be sustained until room temperature in these multilayer heterostructures.

# Results

## IX spectrum

Figure 1 shows the microscopic image and the PL spectrum for different $WSe_2$/$MoS_2$ heterostructure samples. Figure 1a and Fig. 1b shows the schematic of multilayer heterostructure formed by different layers of $MoS_2$ and $WSe_2$. The electrons and holes are separated into the $MoS_2$ layer and the $WSe_2$ layers, respectively, forming the IXs. Figure 1c shows a typical optical image for mL (here m=1 to 4) $WSe_2$/1L-$MoS_2$ sample and Fig. 1d shows the image for 1L-$WSe_2$/mL-$MoS_2$ samples. The 2D layers are first exfoliated mechanically from bulk crystals on polydimethylsiloxane (PDMS) stamps, and then transferred to form the heterostructure by a dry transfer method. The details for sample preparation can be found in the methods section. We used Raman spectroscopy to characterize the sample quality, the layer numbers, and the interlayer coupling strength of the heterostructures (Supplementary Section I Fig. S1-S2).

IX emission and intralayer exciton emission spectrum are shown in Fig. 1e. We note that the IX emission from the heterostructure lies in the infrared range, which is far away from the visible range of all intralayer exciton emission from $WSe_2$ and $MoS_2$. Thus, a pure IX signal can be obtained by using long-pass filters. Figure 1f, g shows the low-temperature PL spectra of IXs from mL-$WSe_2$/1L-$MoS_2$ heterostructure (the sample in Fig. 1c) and 1L-$WSe_2$/mL-$MoS_2$ heterostructure (the sample in Fig. 1d). We can see that, clear IX PL has been observed for all the samples with different spectrum shift. When the number of $WSe_2$ ($MoS_2$) layers is increased, the IXs PL spectra show a redshift up to a hundred nanometers relative to that in the 1L/1L heterostructure case. We prepared several other heterostructure samples and observed similar results as well (more sample images and characterizations are presented in Supplementary Section II and Fig. S3-S9). We anticipate that this redshift is mainly due to the shifts of valance band of mL $WSe_2$ and conduction band of mL-$MoS_2$, as well as the changes of IX binding energy in multilayer heterostructure[28]. This makes IX energy highly tunable with the bandgap engineering[29]. More results and discussions related to this PL redshift in multilayer heterostructure are shown in Supplementary Section III.

## Temperature dependence of IX PL intensity

In order to understand more about the nature of the IX transition and band structure of the stacked heterostructure, we investigate the temperature dependence of PL emissions. Figure 2a, b shows the PL intensity mapping of IXs in mL-WSe$_2$/1L-MoS$_2$ heterostructure at 4.3 K and room temperature, respectively. Here the intralayer exciton emission is blocked by a 1064nm long pass filter and therefore only IX emission is relevant in the PL mapping. The intensity change as a function of temperature is shown in Fig. 2c. Interestingly, for both 1L-WSe$_2$/1L-MoS$_2$ and 2L-WSe$_2$/1L-MoS$_2$ cases, PL intensity decreases as the temperature increases, while PL intensity for 3L-WSe$_2$/1L-MoS$_2$ shows a reversed temperature-dependent behavior. For the former case, the PL temperature dependence is consistent with the case of monolayer MoS$_2$ intralayer exciton[30]. Here, electrons in bright excitons lie in the ground state in the conduction band of MoS$_2$, separated with an upper level with spin flipped. As the temperature is increased, the bright exciton population will decrease since there is a higher probability that phonon can scatter the bright exciton to dark exciton states with electrons lying in a higher energy level. These indicate that the IXs in 1L-WSe$_2$/1L-MoS$_2$ and 2L-WSe$_2$/1L-MoS$_2$ are in a direct K-K transition, similar to intralayer exciton in monolayer MoS$_2$.

In contrast to the 1L-WSe$_2$/1L-MoS$_2$ and 2L-WSe$_2$/1L-MoS$_2$ case, when the number of WSe$_2$ layers is more than two in mL/1L heterostructure (e.g., m=3), the PL intensity of the IXs tends to increase with increasing temperature (Fig. 2c). Similar results are observed in other samples (see Supplementary Section IV Fig. S12). This indicates that the 3L-WSe$_2$/1L-MoS$_2$ heterostructure is a momentum indirect bandgap semiconductor. This comes from the fact that valence band maximum from multilayer WSe$_2$ changes from K to Γ point when the number of layers is increased to three. In this case, the dark state K-Γ transition is in the ground state, while the bright K-K transition has higher energy than the dark transition. Therefore, the PL emission intensity will increase with a higher temperature. We note that these results are consistent with the recent direct measurement of the valance band of 1L-3L WSe$_2$ by the micrometer-scale angle-resolved photoemission spectroscopy (microARPES)[31].

Remarkably, in contrast to the case in mL-WSe$_2$/1L-MoS$_2$ heterostructure, the IX PL intensity in 1L-WSe$_2$/mL-MoS$_2$ (m=1, 3, 4) heterostructure regions is strong at low temperature and it decreases with increasing temperature (Fig. 2d-f). Similar behavior has been observed in other samples up to m = 6 (see Supplementary Section II Fig. S8 and Section IV Fig. S13). These results suggest that 1L-WSe$_2$/mL-MoS$_2$ heterostructure is a direct bandgap semiconductor and the bright state is the ground state, at least for MoS$_2$ up to four layers. To confirm this analysis, we calculate the band structures of 1L-WSe$_2$/mL-MoS$_2$ heterostructure based on the density functional theory (DFT). The calculated results (see Fig. 2g, h and Fig. S14, S15) agree well with the experimental observations.

Additionally, we also note that IX PL intensity in 2L-WSe$_2$/3L-MoS$_2$ heterostructure decreases with increasing temperature, indicating that it is a direct bandgap semiconductor (see Supplementary Section IV Fig. S13). This result is in line with our analysis above that the CBM (VBM) is mainly affected by MoS$_2$ (WSe$_2$) layers with both 3L-MoS$_2$ CBM and 2L-WSe$_2$ VBM located at K(K') point, even though both 2L-WSe$_2$ and 3L-MoS$_2$ are indirect bandgap semiconductor. These results open the door to design the direct bandgap heterostructure semiconductors for future optoelectronics device applications based on the layer engineering of multilayer TMDs.

## Valley polarization for IXs in multilayer heterostructures

Valley polarization has been used to probe the valley index in 2D TMDs. A natural question is: how about the IX valley polarization in our multilayer heterostructures? With optical selection rule (Fig. 3a), the valley is coupled with the PL circularly polarization and therefore, we can characterize the valley properties through circular polarization-resolved PL measurement. Figure 3b shows the corresponding valley degree of polarization (DOP) mapping. Here, we define $DOP = \frac{I_{\sigma^-\sigma^-} - I_{\sigma^-\sigma^+}}{I_{\sigma^-\sigma^-} + I_{\sigma^-\sigma^+}}$, where $I_{\sigma^j\sigma^k}$ represents the PL intensity with $\sigma^j$ excitation and $\sigma^k$ detection, and $\sigma^+(\sigma^-)$ denotes the right (left) circularly polarized light. As shown in Fig. 3b, all heterostructure regions show a high valley DOP. This indicates that the valley optical selection rule is valid in 1L-WSe$_2$/mL-MoS$_2$ heterostructure, i.e. $\sigma^-(\sigma^+)$ IXs emission coupled primarily to K (K') valley. The observation also agrees with the recent report for the K-K transition nature for the interlayer exciton emission in the infrared range for monolayer-monolayer case[8,26]. Similar results are obtained for other 1L-WSe$_2$/mL-MoS$_2$ samples (see Supplementary Section V Fig. S18-S21). Moreover, The IX valley polarization of 1L-WSe$_2$/mL-MoS$_2$ heterostructure is quite robust and can persist up to room temperature, which is desired for valley device applications (Fig. S19).

On the contrary, we find that the DOP for mL-WSe$_2$/1L-MoS$_2$ heterostructure is negligible (See Supplementary Section V Fig. S16). Upon the resonant excitation with $\sigma^-(\sigma^+)$ circularly polarized light (726 nm), the WSe$_2$ excitons are selectively created. Then an ultrafast charge transfer process takes the electrons from the multilayer WSe$_2$ layers to the K (K') valley of monolayer MoS$_2$ layer[32-34], leaving only the holes in the multilayer WSe$_2$ layers[32-34]. Therefore, the vanishing valley polarization in the 2L-WSe$_2$/1L-MoS$_2$ case can only be attributed to the hole interlayer hopping between the WSe$_2$ layers. Additionally, in 3L-WSe$_2$/1L-MoS$_2$ case, the VBM locates at Γ valley which can couple equally to both polarizations. Consequently, the valley polarization of IX vanishes in mL-WSe$_2$/1L-MoS$_2$ (m>1) and 2L-WSe$_2$/3L-MoS$_2$ heterostructures (see Fig. S20). This is in stark difference with 1L-WSe$_2$/mL-MoS$_2$ heterostructure. In the latter case, the VBM at the K (K') valley is mainly contributed by 1L-WSe$_2$, where the orbital mixing of the VBM is largely suppressed. Meanwhile, the interlayer hopping for electrons between mL-MoS$_2$ layers vanishes at K points due to the symmetry of the $d_{z^2}$ orbital[31]. Hence, the valley optical selection rule of IXs in the 1L-WSe$_2$/1L-MoS$_2$ case is maintained for 1L-WSe$_2$/mL-MoS$_2$ heterostructures.

## Enhanced valley lifetime for layer engineered IXs at zero magnetic field

More interestingly, we found that IX in 1L-WSe$_2$/mL-MoS$_2$ (m>1) heterostructure has longer lifetime and valley lifetime than IX in 1L-WSe$_2$/1L-MoS$_2$ heterostructure. Figure 4a-c shows the time-resolved circularly polarized PL of IXs from 1L-WSe$_2$/mL-MoS$_2$ heterostructure regions in Fig. 1d. For the co-polarized emission (i.e., the collected emission polarization is the same as the excitation polarization), the data fits very well with a bi-exponential function. The slowest decay component is up to hundreds of nanoseconds, which is an order of magnitude longer than previously reported results for IX lifetimes[8,35]. All decay components are slower in 1L-WSe$_2$/3L-MoS$_2$ and 1L-WSe$_2$/4L-MoS$_2$ region as compared to the values in the 1L-WSe$_2$/1L-MoS$_2$ region. In contrast, the lifetime of IXs in mL-WSe$_2$/1L-MoS$_2$ heterostructure is decreased with more WSe$_2$ layers (Supplementary Section VI Fig. S22). These results indicate that the number of MoS$_2$ layers can be used to increase the lifetime of IXs.

Based on the time-resolved PL data, we calculate the valley (DOP) lifetime. Figure 4d-f shows the time-resolved DOP at 4.3 K. We fit the data with a single exponential decay function to obtain the valley lifetime. We find that the valley lifetime of IXs depends on the number of $MoS_2$ layers. The DOP decay shows a similar trend with the count decay, i.e., it is getting slower as the number of $MoS_2$ layers is increased. In particular, the valley lifetime increases from 11±0.2 ns in the 1L-$WSe_2$/1L-$MoS_2$ region to 29±0.3 ns in the 1L-$WSe_2$/3L-$MoS_2$ region, and 37±0.5 ns in the 1L-$WSe_2$/4L-$MoS_2$ region (Fig. 4(g)). We get similar results for different samples (Supplementary Section VI Fig. S23-S24). These results indicate that the valley lifetime has been increased by 3-4 times with the increasing number of $MoS_2$ layers.

The layer dependence of the IX lifetime and valley lifetime can be explained by considering that both exciton oscillator strength and electron-hole exchange interaction are suppressed in multilayer heterostructure when the wavefunction overlap between electron and hole is reduced[16]. Similar to the case of multiple quantum well[33], the electron wavefunction spreads more as the number of $MoS_2$ layers is increased. As a result, the reduced electron-hole wavefunction overlap of IXs in 1L-$WSe_2$/mL-$MoS_2$ heterostructure leads to a longer exciton lifetime and valley lifetime.

## Magnetic field enhanced valley polarization and valley lifetime

With applied magnetic field, we found that the valley polarization and valley lifetime can be further enhanced for IXs in 1L-$WSe_2$/mL-$MoS_2$. Fig. 5a shows co-polarized and cross-polarized PL spectra. One can clearly see the enlarged difference in spectrum between the two cases, representing an enhanced valley polarization. To examine the magnetic field dependence in more detail, we plot DOP under different magnetic field, as shown in Fig. 5b. It shows that the amplitude of DOP has increased dramatically from ~0.1-0.2 to ~ 0.7-0.8 with applying a tiny magnetic field. This is the case for all the three situations of IXs in 1L-$WSe_2$/mL-$MoS_2$ (m=1, 3, 4). This implies that the origin of increased DOP is similar in three cases. Since we are exciting resonantly with $WSe_2$, this shows the suppression of valley relaxation in $WSe_2$ excitons with magnetic fields. For IXs in 1L-$WSe_2$/4L-$MoS_2$, a prolonged valley lifetime has been shown in Fig. 5c and 5d as well, which originated from $WSe_2$ dark excitons.

More interestingly, it is worth noting that the valley relaxation is slow in 1L-$WSe_2$/4L-$MoS_2$ such that the valley mixing rate is comparable with the PL decay lifetime. This leads to the fact that the cross-polarized PL count shows an initial count increase, rather than an exponential decay, as shown in Fig. 5e. To extract the information for this valley mixing mechanism, we ramp up the magnetic field. One can clearly see the suppression of the cross-polarized PL counts with the increase of magnetic field, which comes from the suppression of valley mixing. One can give an estimation of the valley mixing rate 5 ±1 ns in this case, by subtracting the count rate in the magnetic field from the zero field case. This represents the first direct observation of valley relaxation in the time domain from PL emission, thanks to the prolonged exciton lifetimes.

## Conclusion

In summary, we studied the IXs from multilayer $WSe_2$/$MoS_2$ heterostructures. We demonstrated that the layer number of $WSe_2$ and $MoS_2$ offers an additional degree of freedom to modulate the emission spectra, as well as enhancing exciton lifetime, the valley polarization, and the valley lifetime of IXs. The 1L-$WSe_2$/mL-$MoS_2$ heterostructure preserves the direct bandgap at least for $MoS_2$ up to four layers. These results demonstrate that novel phenomena can be achieved by layer engineering of multilayer 2D semiconductor materials, which offers a systematic approach readily extended to other

2D materials. The layer engineering combines with the twist angle in vdW heterostructure will bring many opportunities for fundamental research and optoelectronics device applications.

## Methods

**Heterostructures fabrication:** The monolayer and few-layer $MoS_2$, monolayer and few-layer $WSe_2$ are first exfoliated mechanically from bulk crystals on polydimethylsiloxane (PDMS) stamps. The alignment between $MoS_2$ and $WSe_2$ layers is done by first choosing these samples that have two sharp edges with a 120° angle. We then used the dry transfer method to stack $MoS_2$ and $WSe_2$ samples by aligning the edge of the top and bottom samples onto an ultralow doping Si substrate that is covered by 285nm $SiO_2$, to form the $WSe_2/MoS_2$ and $hBN/WSe_2/MoS_2/hBN$ heterostructures (HS). Finally, the HS samples were annealed under ultrahigh vacuum (around $10^{-6}$ mbar) at 200 °C for 3 hours.

### Optical measurements

**PL measurement:** A homemade confocal microscope is used to perform the polarization and spatially resolved PL spectroscopy. The polarization state of the excitation and the detection are controlled by a combination of polarizers, half-wave plates, and quarter-wave plates installed on the excitation and detection. The PL spectra were obtained by a spectrometer (Andor Shamrock) with a CCD detector. Unless otherwise stated, a laser with 726 nm was used as the excitation source at low temperatures. A 50x objective lens with spot size 1 μm (NA=0.65) is used to collect the signal. The sample temperature is controlled using a cryostat (Montana Instruments). For pulse experiments, a 726 nm diode laser (FWHM<80 ps, Max. rep. rate is 80MHz) is used for exciting the sample. The intensity map is obtained by detecting the emission using a superconducting single-photon detector while scanning the excitation position on the sample by using a galvo system. In all of the optical measurements, the CW laser power is around 100 μW.

**Raman measurement:** The Raman spectra were measured at room temperature with the T64000 Raman system, equipped with a liquid-nitrogen-cooled CCD. A 50x objective lens (NA=0.95) is used to collect the signal. The excitation wavelength for Raman spectra measurement is 532 nm.

### Theoretical calculations

We performed first-principles calculations using the Vienna ab initio simulation package[12,36] with the projector augmented wave method. The Perdew-Burke-Ernzerhof-type[37] generalized gradient approximation was used for the exchange-correlation functional. The heterostructures were constructed by taking the average of the experimental lattice constant of bulk materials (the lattice constants for $MoS_2$ is a=b=3.168 Å[38] and $WSe_2$ is a=b=3.282 Å[39]. We set the interlayer distance to be 6.67 Å between $MoS_2$ and $WSe_2$ initially and then relax the interlayer distance and atomic coordinates. The cutoff energy was set to 500 eV, and a 16 × 16 × 1 Γ-centered k-point mesh was used for the Brillouin zone sampling. The convergence criteria for the energy and force were set to be $10^{-5}$ eV and 0.01 eV/Å, respectively. A vacuum layer with a thickness of 20 Å was taken to avoid artificial interactions between periodic images. The van der Waals corrections were taken into account by the approach of Dion et al[40]. To get a more accurate energy gap, we also used the more sophisticated Heyd−Scuseria−Ernzerhof hybrid functional method (HSE06) for the exchange-correlation potential[41] to calculate the band structure for the 1L/1L HS.

# Figures

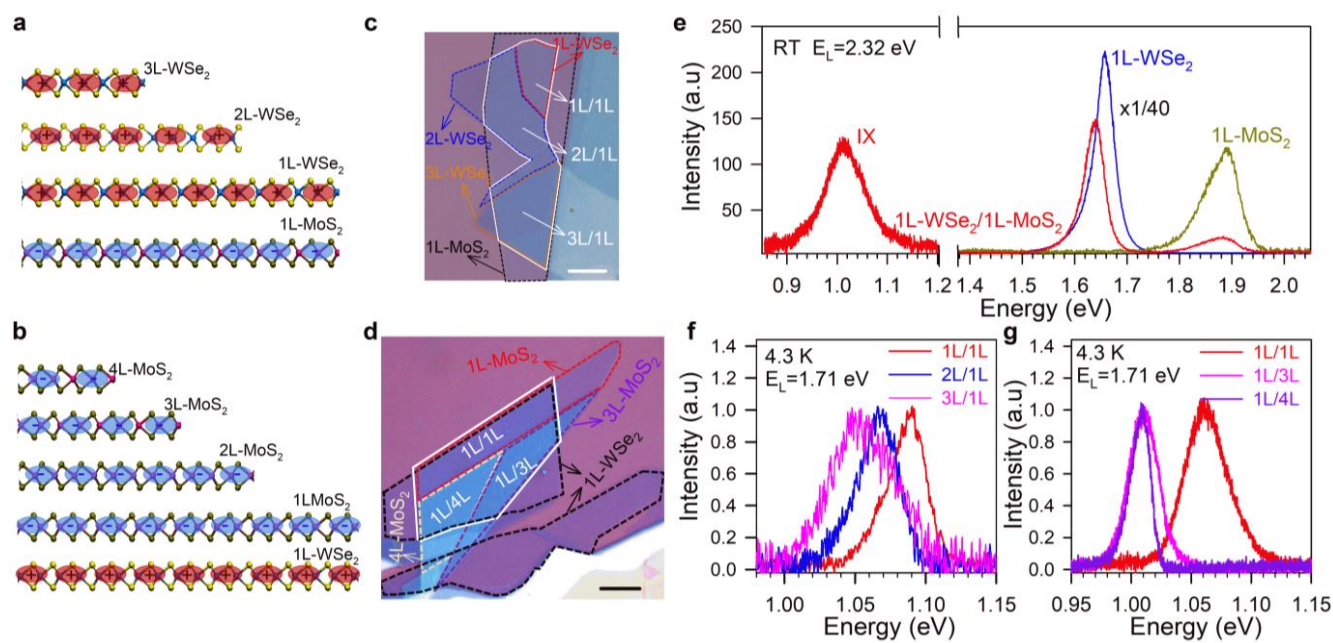

**Fig. 1 | Observation of interlayer excitons in mL-WSe$_2$/1L-MoS$_2$ and 1L-WSe$_2$/mL-MoS$_2$ heterostructures (HS). a, b,** A schematic of the interlayer excitons in mL-WSe$_2$/1L-MoS$_2$ and 1L-WSe$_2$/mL-MoS$_2$ HS, respectively. The electrons and holes are separated in MoS$_2$ and WSe$_2$ layers. **c, d,** The optical microscope image of mL-WSe$_2$/1L MoS$_2$ (labelled as mL/1L, m=1 to 3) sample (S1) and 1L-WSe$_2$/mL-MoS$_2$ (labelled as 1L/mL, m=1, 3 and 4) sample (S2). WSe$_2$ and MoS$_2$ with different layers are marked with dish lines of different colours. The heterostructure regions are marked with solid white lines. The scale bar is 10 μm. **e,** The PL spectra of intralayer excitons in monolayer WSe$_2$, MoS$_2$, and interlayer excitons in 1L-WSe$_2$/1L-MoS$_2$ heterostructure sample at room temperature (RT). **f, g,** The PL spectra of interlayer excitons in mL-WSe$_2$/1L-MoS$_2$ heterostructure and 1L-WSe$_2$/mL-MoS$_2$ heterostructure at low temperature.

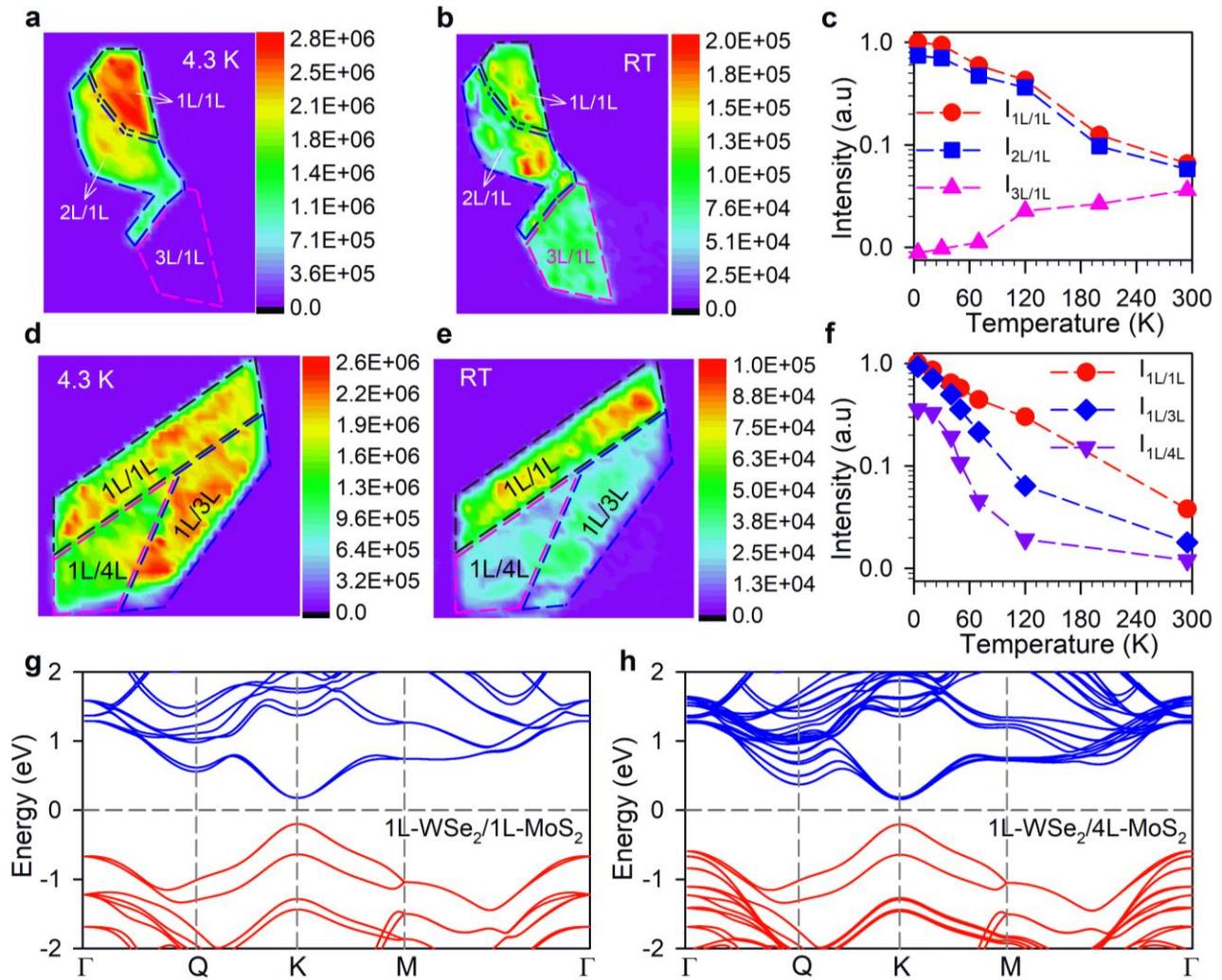

**Fig. 2 | Temperature dependence of IXs in mL-WSe$_2$/1L-MoS$_2$ and 1L-WSe$_2$/mL-MoS$_2$ HS. a, b,** The PL intensity map of the IXs in mL-WSe$_2$/1L-MoS$_2$ (mL/1L, m=1 to 3) HS at 4.3 K and room temperature, respectively. **c,** The PL intensity of IXs in mL-WSe$_2$/1L-MoS$_2$ as a function of temperature. **d, e,** The PL intensity map of the IXs in 1L-WSe$_2$/mL-MoS$_2$ (mL/1L, m=1, 3 and 4) HS at 4.3 K and room temperature, respectively. A 1064 nm long pass is used here to ensure that only IX signal can be detected. **f,** The PL intensity of IXs in 1L-WSe$_2$/mL-MoS$_2$ as a function of temperature. **g, h,** The calculated electronic energy structure of 1L-WSe$_2$/1L-MoS$_2$ heterostructure and 1L-WSe$_2$/4L-MoS$_2$ heterostructure, respectively.

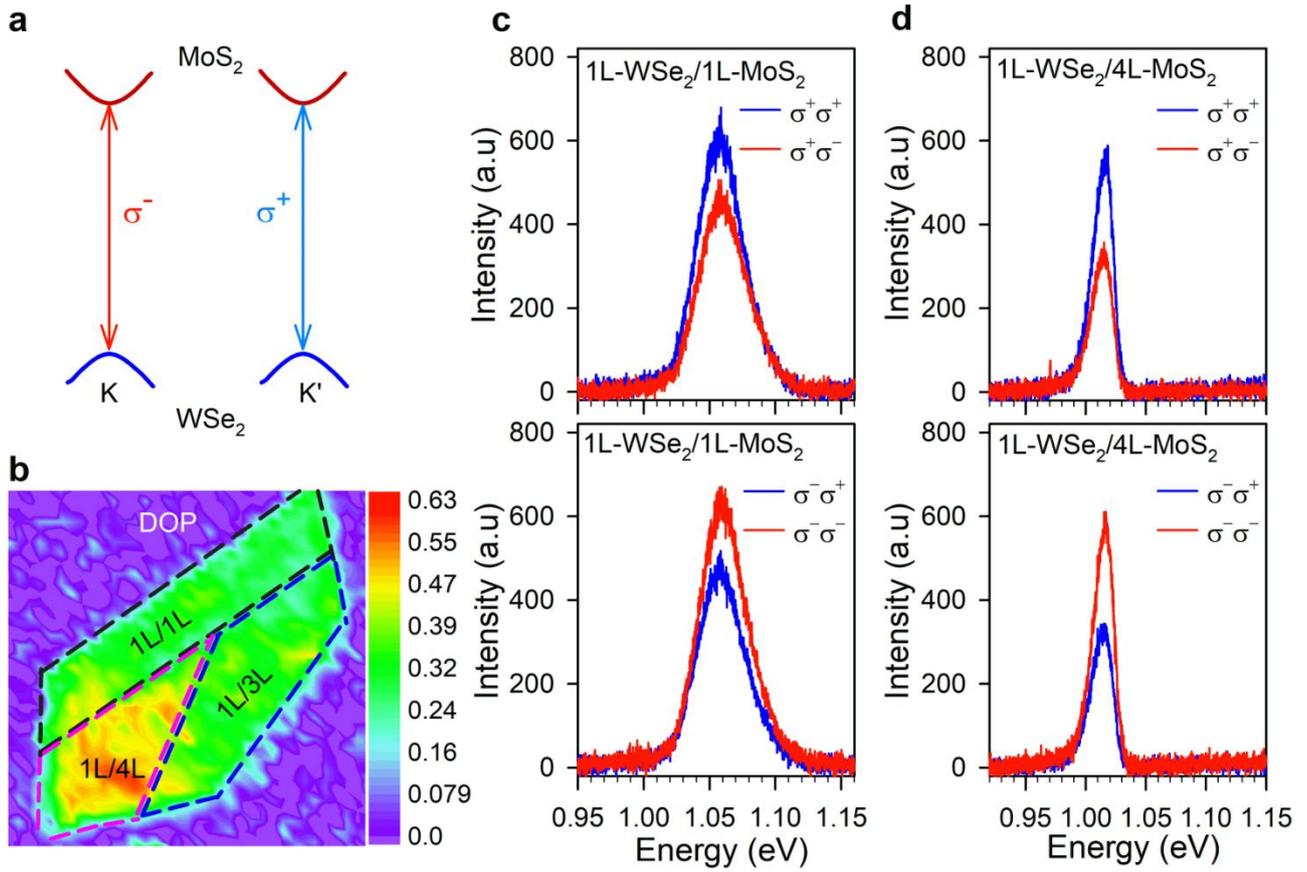

**Fig. 3 | Circularly polarized PL spectra of IXs in 1L-WSe$_2$/mL-MoS$_2$ HS. a**, A simplified schematic of valley optical selection rule for IXs in K and K′ valley. The electrons and holes are located at K (K′) point of MoS$_2$ conduction band and WSe$_2$ valance band, respectively. **b**, The valley degree of polarization (DOP) mapping of IXs in the multilayer HS calculated from the polarized PL intensity mapping measurement results. **c-d**, The circularly polarized PL spectra of IXs from 1L-WSe$_2$/1L-MoS$_2$ and 1L-WSe$_2$/4L-MoS$_2$ heterostructure regions, respectively. Here σ$^i$σ$^j$ represents excitation with σ$^i$ and detection with σ$^j$ circular polarization.

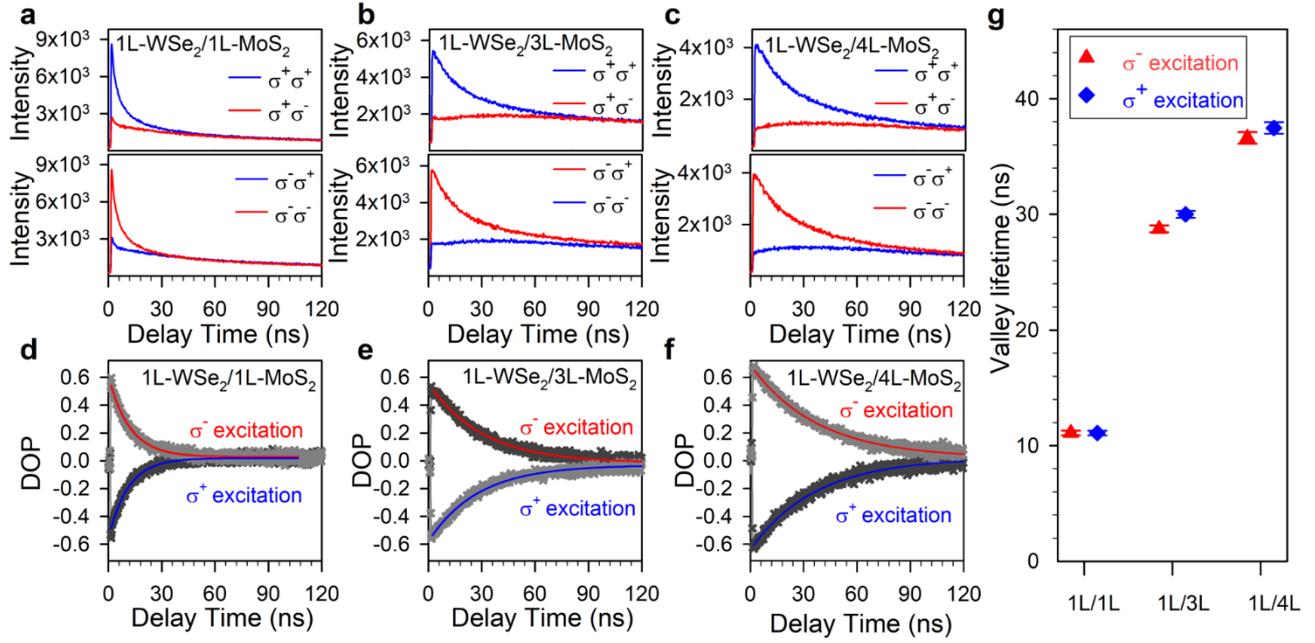

**Fig. 4 | Layer-engineered depolarization lifetime of IX in 1L-WSe$_2$/mL-MoS$_2$ HS**. **a-c**, Time-resolved circularly polarized PL of IXs from 1L-WSe$_2$/1L-MoS$_2$, 1L-WSe$_2$/3L-MoS$_2$, 1L-WSe$_2$/4L-MoS$_2$ HS regions, respectively. The IXs lifetime of a few hundred ns was observed. **d-f**, The time-resolved valley DOP of IXs obtained from the measured time-resolved circularly polarized PL in (a-b). **g**, The valley DOP lifetimes under left and right circularly polarized light excitation.

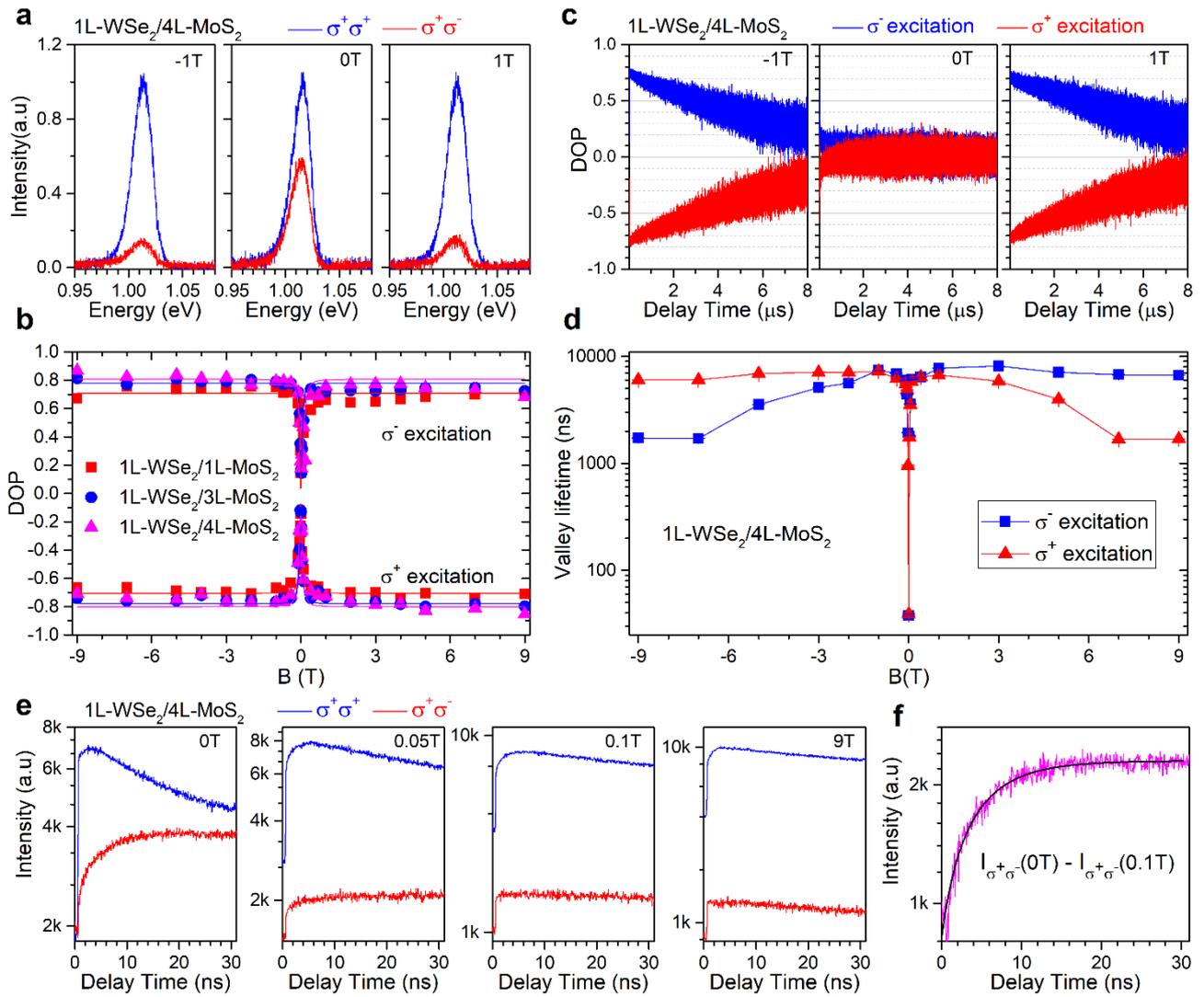

**Fig. 5 | Suppression of valley relaxation under magnetic field for IXs in 1L-WSe$_2$/mL-MoS$_2$ HS. a,** Valley polarization for IXs in 1L-WSe$_2$/4L-MoS$_2$ under three magnetic fields -1T, 0T and 1T. An enhanced valley polarization under magnetic field has been observed. **b,** Magnetic dependence of degree of polarization (DOP). Valley mixing has been suppressed for all m=1, 3, 4 cases. **c,** Time-resolved DOP for IXs in 1L-WSe$_2$/4L-MoS$_2$ at -1T, 0T and 1T. An enhanced valley polarization lifetime has been observed under magnetic field. **d,** The magnetic field dependence of valley DOP lifetime under left and right circularly polarized light excitation, respectively. **e,** Time-resolved photon emission at different magnetic field. **f,** The magnetic field suppressed valley relaxation characterization.

# Supplementary Information for "Layer engineered interlayer excitons"

## Section 0. Second harmonic generation (SHG) for the alignment of the angle.

In the sample preparation, we have aligned the angle to zero or 60 degree, which will give strong exciton emission. The angle is confirmed with angle dependent SHG measurement as shown in Fig. S0.

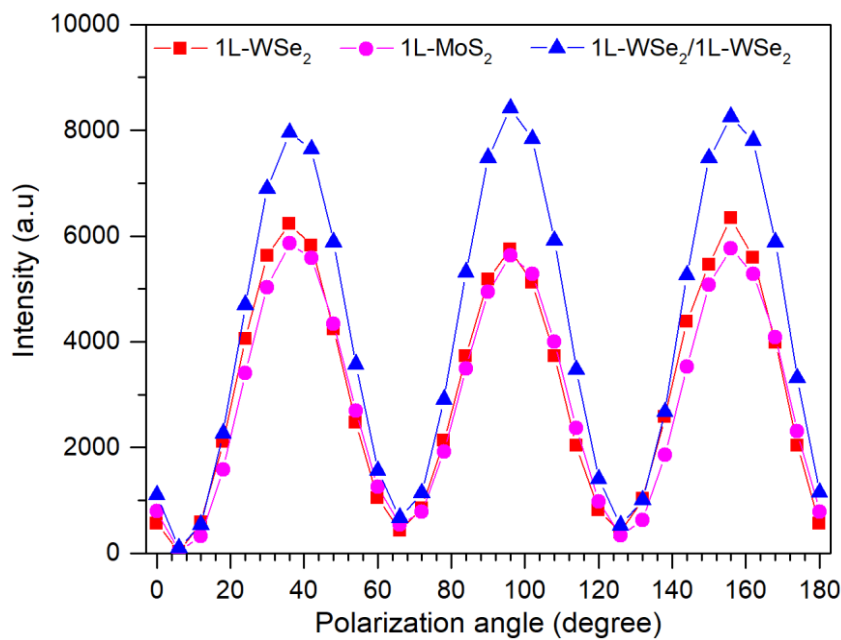

**Fig. S0. The second harmonic generation signal as a function of polarization angle.**

## Section I. Raman spectroscopy of the mL-WSe$_2$, mL-MoS$_2$, the mL-WSe$_2$/1L-MoS$_2$ heterostructures and the 1L-WSe$_2$/mL-MoS$_2$ heterostructures.

To understand the role of the interlayer coupling between HS interface, we measured the ultralow to high-frequency Raman spectra of 1L to 3L WSe$_2$ and the mL-WSe$_2$/1L-MoS$_2$ heterostructures (HS) region (Fig. S1). In addition, we have measured 1L to 4L MoS$_2$ and 1L-WSe$_2$/mL-MoS$_2$ HS (Fig. S2). In the high-frequency range (above 100 cm$^{-1}$), in addition to the $E^1_{2g}$ and $A_{1g}$ modes from WSe$_2$ and MoS$_2$, two more feature modes are observed in 1L-WSe$_2$/1L-MoS$_2$ HS region. These two modes are the Raman-inactive $E^2_{2g}$ mode (~285 cm$^{-1}$) in monolayer MoS$_2$ and the $B^1_{2g}$ (~ 310 cm$^{-1}$) mode in monolayer WSe$_2$. Both are activated in 1L/1L HS, owing to the reduced symmetry of the HS [1,2]. For the same reason, more feature modes are observed in mL-WSe$_2$/1L-MoS$_2$ HS than in multilayer WSe$_2$. In the ultralow frequency range (below 50 cm$^{-1}$), we found a series of shear (S) and layer breathing (LB) modes. Their mode frequencies are layer dependent for layered 2D materials [3,4]. We note that the S modes in mL-WSe$_2$/1L-MoS$_2$ HS are almost the same with mL WSe$_2$, suggesting that these modes are originated from mL WSe$_2$. On the other hand, the LB modes in mL-WSe$_2$/1L-MoS$_2$ HS behaves like (1+m) L WSe$_2$. For example, in the 1L/1L HS region, a new LB mode is observed with a frequency slightly larger than that in bilayer WSe$_2$. The same case also can be found in the 1L-WSe$_2$/mL-MoS$_2$ HS. A similar case has also been found in the twisted multilayer graphene[5].

The interlayer coupling strength can be represented by the coupling force constant per unit area, $\alpha$. The value of $\alpha$ can be obtained directly from the frequency of the LB mode in the bilayer 2D MX$_2$ (M=Mo, W; X=S, Se) materials by the formula[3] $\alpha = 2\mu\left[\pi c \omega_{LB}(2)\right]^2$, where $c$ is speed of light, $\alpha$ is the force constant per unit area between the adjacent layers, and $\mu = m_M + 2m_X$ with $m_M$ and $m_X$ are the mass of the M and X component, respectively [1,3]. The frequencies of the LB modes are around 41 cm$^{-1}$, 27.6 cm$^{-1}$, and 31 cm$^{-1}$ in the bilayer MoS$_2$, WSe$_2$, and HS, respectively. Therefore, the interlayer coupling strength in the multilayer WSe$_2$ is around 96.8% of the multilayer MoS$_2$, while in the HS, it is approximately 94.5% of the multilayer WSe$_2$. Given the force constant, we can calculate the frequencies of the LB modes in the multilayer WSe$_2$ by using the linear chain model [3,4]. Fig. S1(b) shows the calculated frequencies of the LB modes in 1L to 5L WSe$_2$ as well as the experimental results of the LB modes in 1L to 4L WSe$_2$ and its HS region. As can be seen from this figure, the theoretical results agree well with the experimental results.

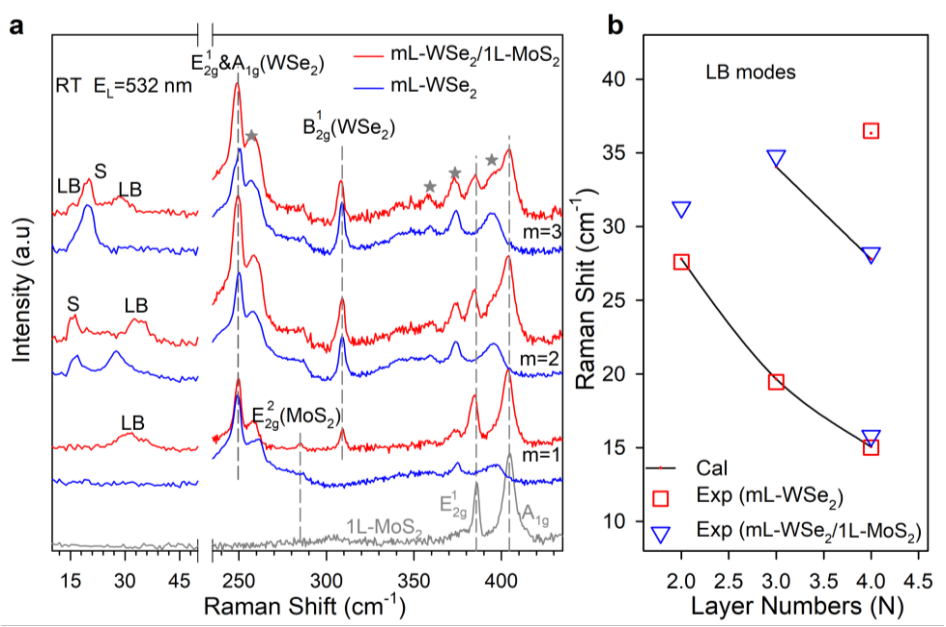

**Fig. S1. The Raman spectra of the m-layer WSe$_2$, the monolayer MoS$_2$, and the mL-WSe$_2$/1L-MoS$_2$ heterostructures. a,** The Raman spectra of the mL-WSe$_2$, the monolayer MoS$_2$, and the mL-WSe$_2$/1L-MoS$_2$ (m=1 to 3) heterostructures (HS) (sample S1). The excitation wavelength is 532 nm. The ultralow frequency shear (S) and layer breathing (LB) modes can be used to identify the layer numbers of WSe$_2$ and HS. **b,** The calculated layer breathing (LB) modes and the experimental results of the mL-WSe$_2$ and the mL-WSe$_2$/1L-MoS$_2$ (m=1 to 3) HS.

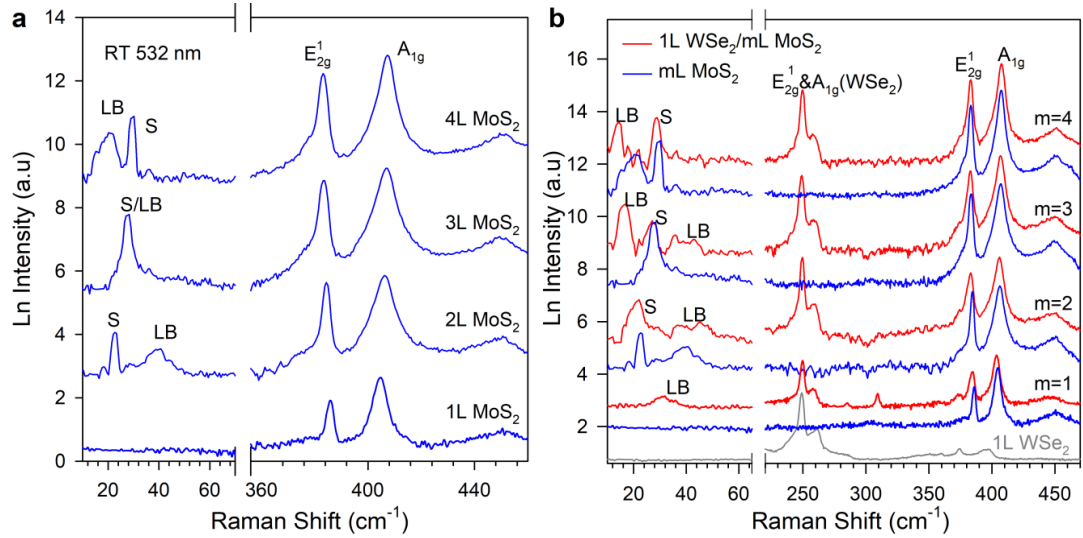

**Fig. S2 | The Raman spectra of the m-layer MoS$_2$, the monolayer WSe$_2$, and the 1L-WSe$_2$/mL-MoS$_2$ heterostructures. a,** The Raman spectra of 1L to 4L-MoS$_2$ at room temperature. **b,** The Raman spectra of the mL-MoS$_2$ and the 1L-WSe$_2$/mL-MoS$_2$ heterostructures (m=1 to 4) at room temperature. The excitation wavelength is 532 nm. The ultralow frequency shear (S) and layer breathing (LB) modes are used to identify the layer numbers of MoS$_2$ and HS. Here the measured spectra are from sample 2 in main text and sample presented in Fig. S8.

## Section II. More multilayer mL-WSe$_2$/1L-MoS$_2$ and 1L-WSe$_2$/mL-MoS$_2$ heterostructures (HS) samples

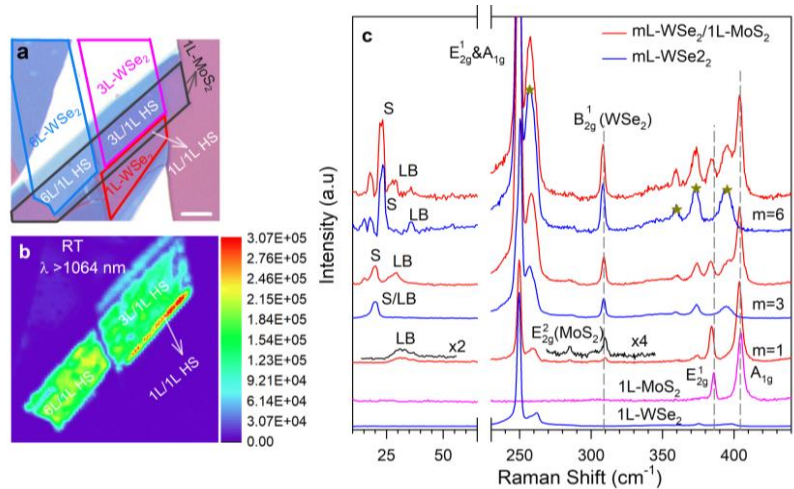

**Fig. S3 | The optical microscope image and the Raman spectra of the mL-WSe$_2$/1L-MoS$_2$ HS. a,** The optical microscope image of the mL-WSe$_2$/1L-MoS$_2$ (m=1, 3, 6) HS (sample S3). The scale bar is 5 μm. **b,** The PL intensity map of IXs in the HS of sample S3 at room temperature. A 1064 nm long pass (LP1064) and 726 nm excitation wavelength are used here. **c,** The Raman spectra of the mL-WSe$_2$ and the mL-WSe$_2$/1L-MoS$_2$ HS at room temperature.

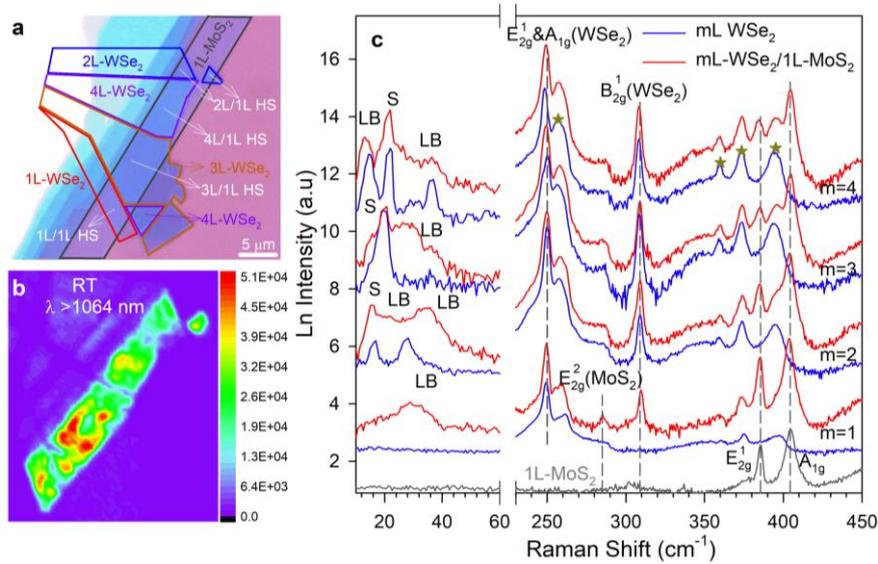

**Fig. S4 | The optical microscope image and the Raman spectra of the mL-WSe$_2$/1L-MoS$_2$ HS. a,** The optical microscope image of the mL-WSe$_2$/1L-MoS$_2$ (m=1 to 4) HS (sample S4). **b,** The PL intensity map of IXs in the HS of sample S4 at room temperature. A 1064 nm long pass (LP1064) and 726 nm excitation wavelength are used here. **c,** The Raman spectra of the mL-WSe$_2$ and the mL-WSe$_2$/1L-MoS$_2$ HS at room temperature.

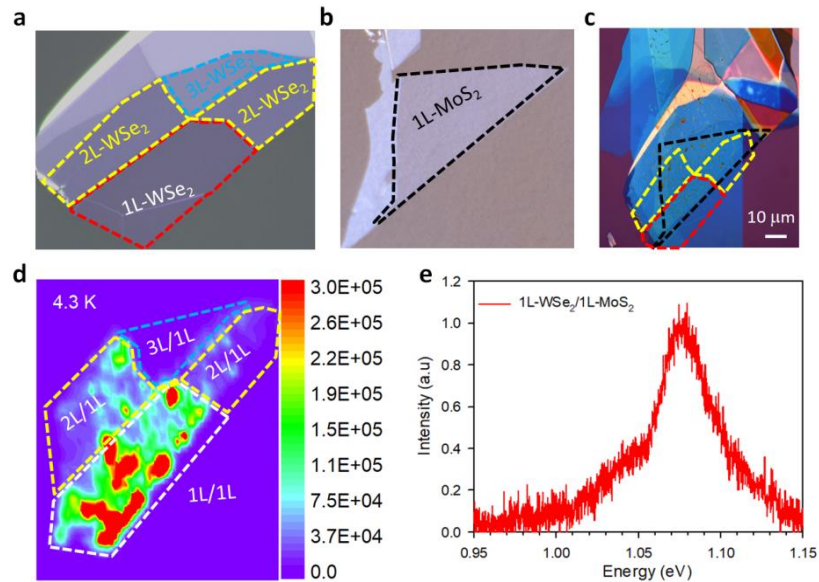

**Fig. S5 | The optical microscope image and the PL spectra of IXs in the hBN/mL-WSe$_2$/1L-MoS$_2$/hBN HS. a-c,** The optical microscope image of the mL-WSe$_2$, the 1L-MoS$_2$, and the mL-WSe$_2$/1L-MoS$_2$ (m=1 to 3) HS encapsulated with the hBN, respectively. **d,** The PL intensity map of IXs at 4.6 K. **e,** The PL spectra of IXs in 1L-WSe$_2$/1L-MoS$_2$ HS encapsulated with the hBN. The linewidth is close to that in the HS without the hBN encapsulation.

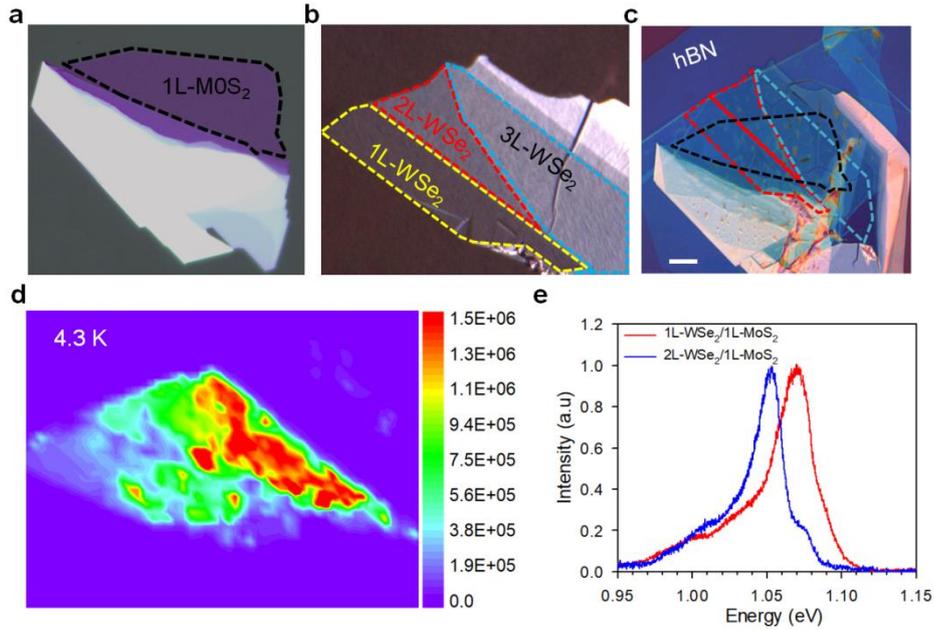

**Fig. S6 | The optical microscope image and the PL spectra of IXs in the hBN/mL-WSe$_2$/1L-MoS$_2$/hBN HS**. **a-c,** The optical microscope image of the mL-WSe$_2$, the 1L-MoS$_2$, and the mL-WSe$_2$/1L-MoS$_2$ (m=1 to 3) HS encapsulated with the hBN, respectively. **d,** The PL intensity map of IXs at 4.6 K. **e,** The normalized PL spectra of IXs in the 1L-WSe$_2$/1L-MoS$_2$ HS and the 2L-WSe$_2$/1L-MoS$_2$ HS encapsulated with the hBN. The linewidth is close to that in the HS without the hBN encapsulation.

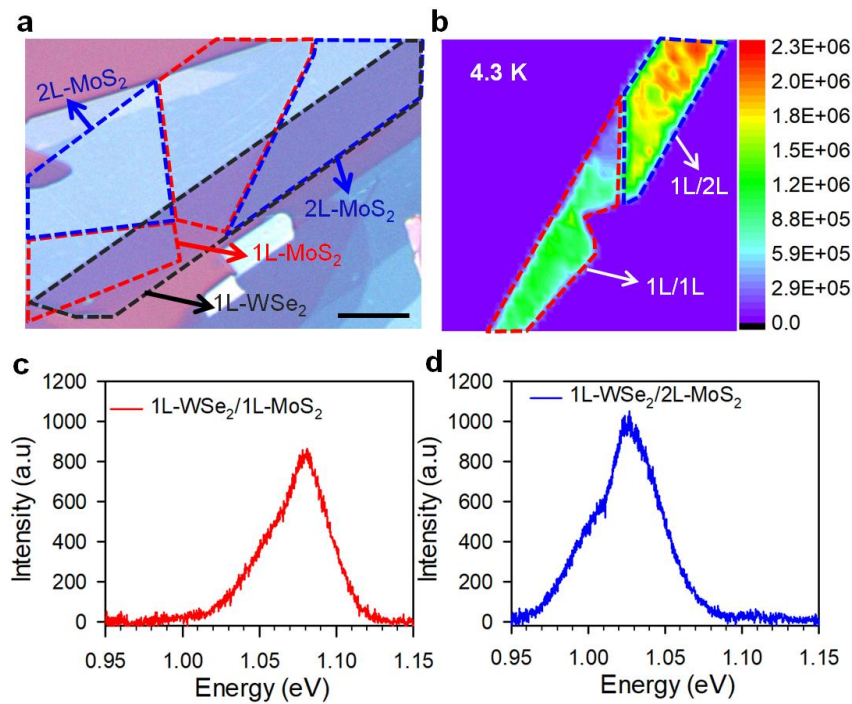

**Fig. S7 | The optical microscope image and the PL spectra of IXs in the 1L-WSe$_2$/mL-MoS$_2$ HS. a,** The optical microscope image of the 1L-WSe$_2$/1L-MoS$_2$ and the 1L-WSe$_2$/2L-MoS$_2$ HS (sample S5). The scale bar is 10 μm. **b,** The PL intensity mapping of IXs at 4.3 K. **c, d,** The PL spectra of IXs from the 1L-WSe$_2$/1L-MoS$_2$ and the 1L-WSe$_2$/2L-MoS$_2$ HS at 4.3 K.

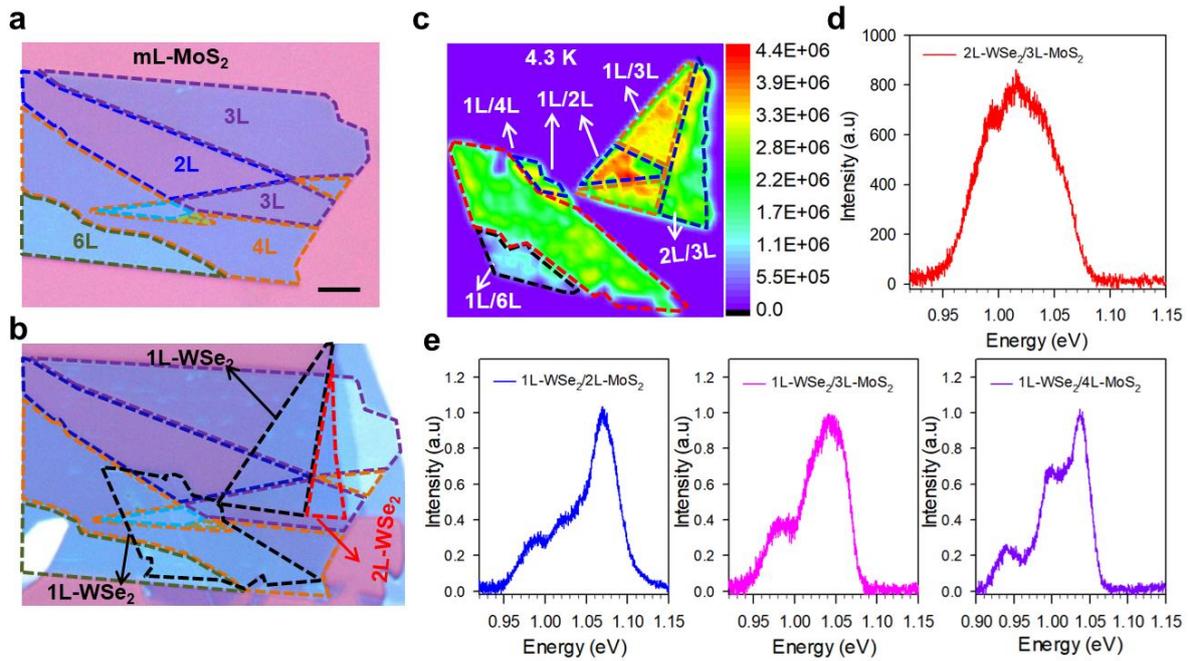

**Fig. S8 | The optical microscope image and the PL spectra of IXs in the 1L-WSe$_2$/mL-MoS$_2$ HS. a,** The optical microscope image of the mL-MoS$_2$ (m=2 to 6). The scale bar is 10 μm. **b,** The optical microscope image of the 1L-WSe$_2$/mL-MoS$_2$ and the 2L-WSe$_2$/3L-MoS$_2$ HS (sample S6). **c,** The PL intensity mapping of IXs at 4.3 K. **d, e,** The PL spectra of IXs from the 2L-WSe$_2$/3L-MoS$_2$, the 1L-WSe$_2$/2L-MoS$_2$, the 1L-WSe$_2$/3L-MoS$_2$, and the 1L-WSe$_2$/4L-MoS$_2$ HS regions at 4.3 K, respectively. The multiple peaks of IXs may result from the Moiré superlattice. There is no correlation between the number of peaks and the number of layers.

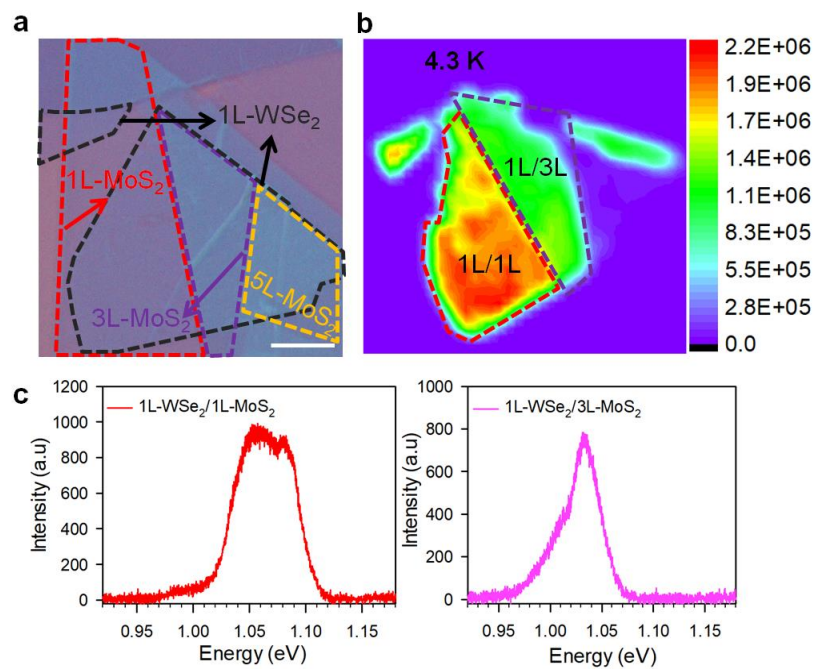

**Fig. S9 | The optical microscope image and the PL spectra of IXs in the 1L-WSe$_2$/mL-MoS$_2$ HS. a,** The optical microscope image of the 1L-WSe$_2$/mL-MoS$_2$ HS (S7). **b,** The PL intensity mapping of IXs at 4.3 K. **c,** The PL spectra of IXs from the 1L-WSe$_2$/1L-MoS$_2$, and the 1L-WSe$_2$/3L-MoS$_2$ HS regions at 4.3 K, respectively.

**Section III. Analysis of the PL redshift of IXs in the mL-WSe$_2$/1L-MoS$_2$ and the 1L-WSe$_2$/mL-MoS$_2$ HS**

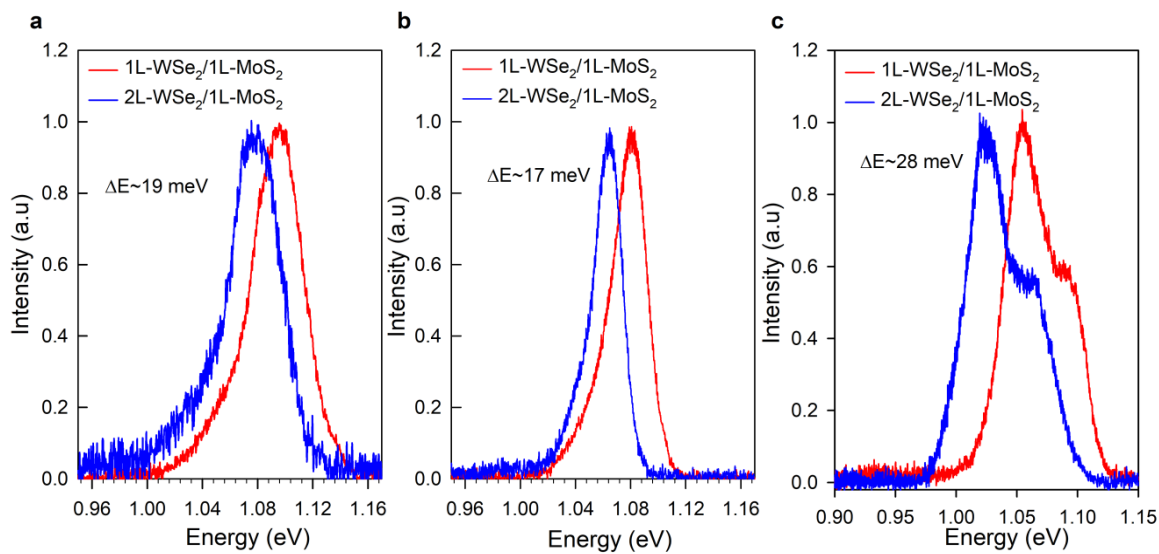

**Fig. S10 | Layer engineered interlayer emission energies in the mL-WSe$_2$/1L-MoS$_2$ HS. a-c,** Comparison between the normalized PL spectra of 1L-WSe$_2$/1L-MoS$_2$ HS vs. 2L-WSe$_2$/1L-MoS$_2$ HS, showing the energy difference of the IXs emission from three different HS samples. The comparison is done for the 1L-WSe$_2$/1L-MoS$_2$ HS and the 2L-WSe$_2$/1L-MoS$_2$ HS regions from the same sample.

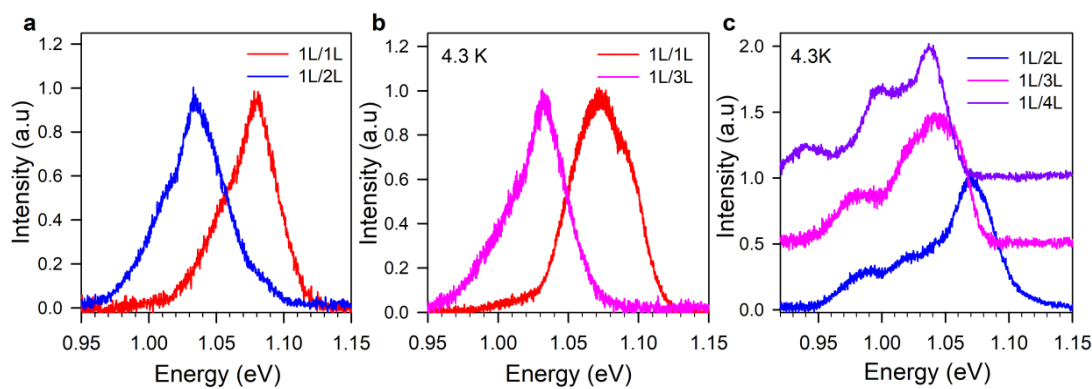

**Fig. S11 | Layer engineered interlayer emission energies in the 1L-WSe$_2$/mL-MoS$_2$ HS. a-c**, The normalized PL spectra of the 1L-WSe$_2$/1L-MoS$_2$ HS and the 1L-WSe$_2$/2L-MoS$_2$ HS (sample S5) (a), the 1L-WSe$_2$/1L-MoS$_2$ and 1L-WSe$_2$/3L-MoS$_2$ HS (sample S7) (b), and the 1L-WSe$_2$/2L-MoS$_2$, 1L-WSe$_2$/3L-MoS$_2$, and 1L-WSe$_2$/4L-MoS$_2$ HS (sample S6) (c).

The IXs PL energy redshifts as the number of WSe$_2$ (MoS$_2$) layer increases in multilayer HS. This can be attributed to the combined effect of the WSe$_2$ (MoS$_2$) VBM (CBM) band shift and the reduced IXs binding energy. The increased effective electron-hole distance in the multilayer HS relative to the 1L/1L HS will reduce the IXs binding energy [6,7], i.e., the IXs binding energy decreases from ~283 meV in the normal 1L-WSe$_2$/1L-MoS$_2$ HS to ~233 (238) meV in one vacuum (or hBN) layer intercalated 1L-WSe$_2$/1L-MoS$_2$ HS[6]. If we only consider the binding energy change, this reduced value will lead to a blue shift of the IXs PL peaks in the mL-WSe$_2$/1L-MoS$_2$ and 1L-WSe$_2$/mL-MoS$_2$ HS, which is opposite to our experimental results. Therefore, the observed red-shift is mainly due to the band shift. Based on the measured results and the estimated binding energy, the valance band K point energy shift from the monolayer $WSe_2$ to the bilayer $WSe_2$ is estimated to be $47 \pm 6$ meV, which is consistent with the value ($50 \pm 10$ meV) measured by angle-resolved photoemission spectroscopy (ARPES)[8].

## Section IV. Temperature dependence and power dependence of IXs in the mL-$WSe_2$/1L-$MoS_2$ HS.

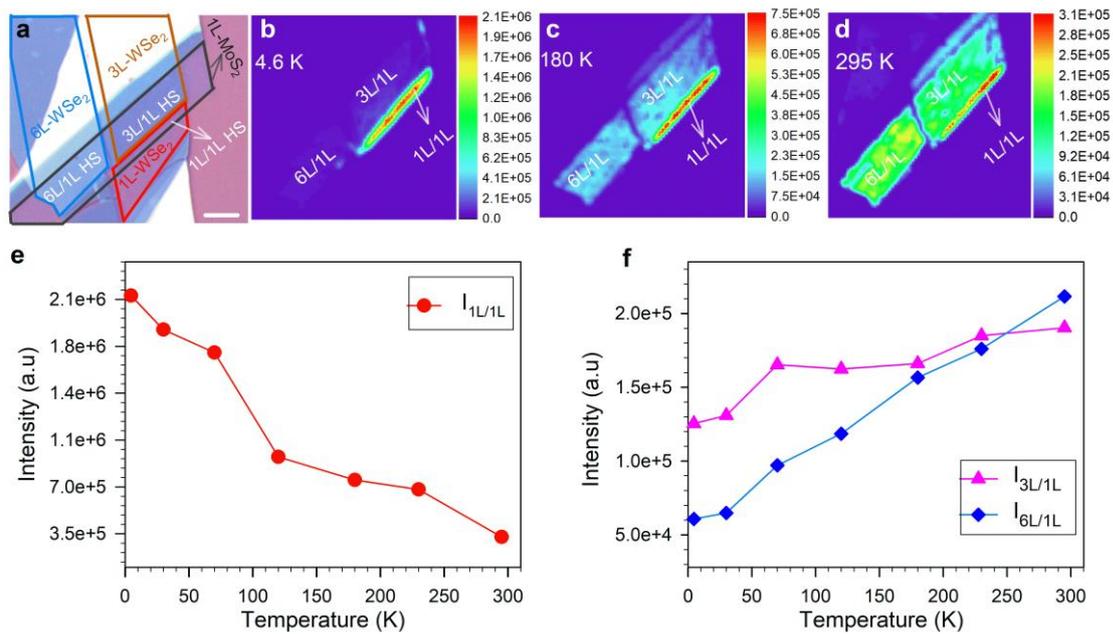

**Fig. S12 | Temperature dependence of IXs in the mL-$WSe_2$/1L-$MoS_2$ HS. a,** The optical microscope image of the mL/1L (m=1, 3, 6) HS (S3), the scale bar is 10 μm. **b-d,** The PL intensity map of IXs at 4.6 K, 180 K, and 295 K, respectively. **e, f,** The PL intensities of mL-$WSe_2$/1L-$MoS_2$ HS as a function of temperature.

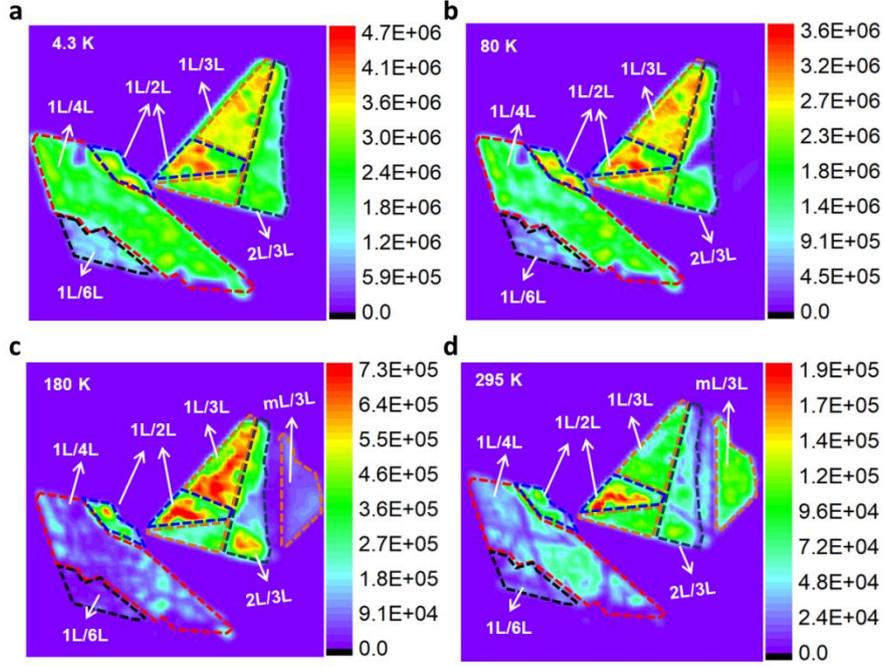

**Fig. S13 | Temperature dependence of IXs in the multilayer WSe$_2$/MoS$_2$ HS. a-d,** The PL intensity map of IXs at 4.6 K, 80 K, 180 K, and 295 K, respectively. With the temperature increase from 4.3 K to room temperature, the PL intensities in 1L-WSe$_2$/mL-MoS$_2$ (m=2-6) HS and 2L-WSe$_2$/3L-MoS$_2$ regions decreases, while the PL intensities in mL-WSe$_2$/3L-MoS$_2$ (m>6) regions increases.

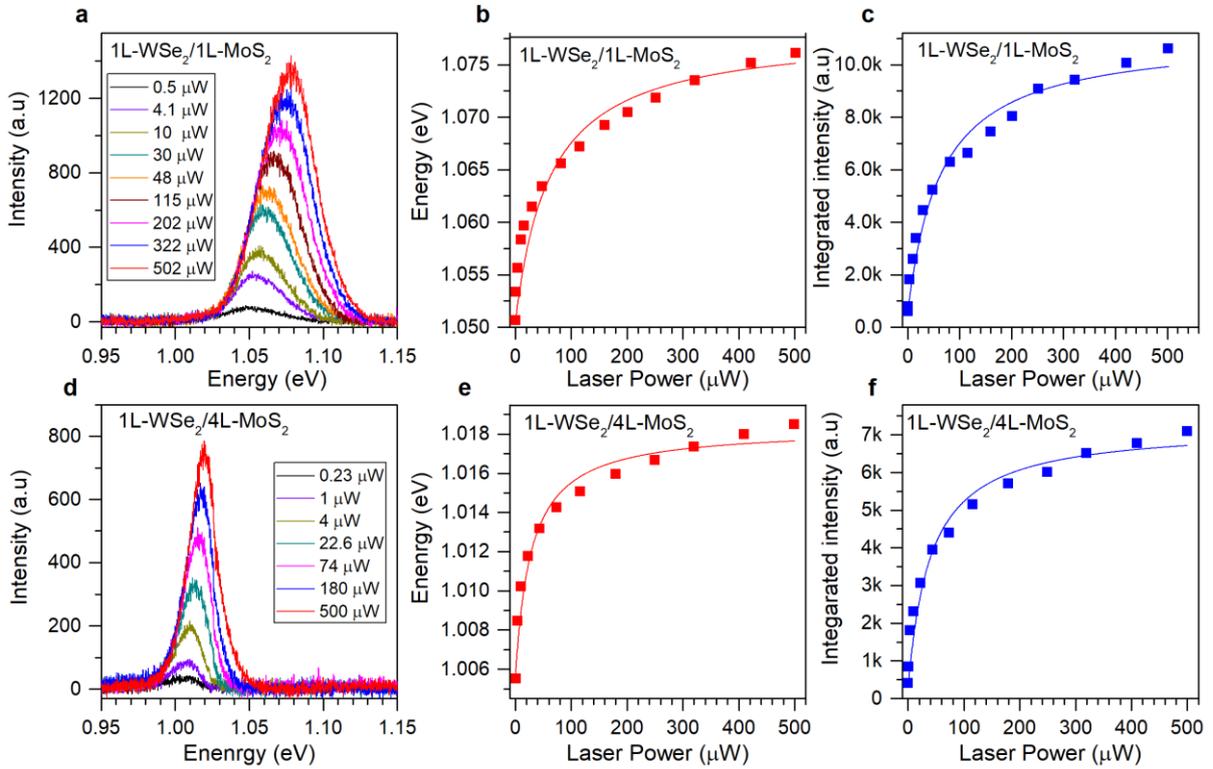

**Fig. S14 | Power dependence of IXs in the 1L-WSe$_2$/mL-MoS$_2$ HS. a,** The power dependence of PL spectra of IXs in 1L-WSe$_2$/1L-MoS$_2$ HS. **b, c,** The PL peak energy and integrated intensity as a function of

laser power in 1L-WSe$_2$/1L-MoS$_2$ HS. **d,** The power dependence of PL spectra of IXs in 1L-WSe$_2$/4L-MoS$_2$ HS. **b, c,** The PL peak energy and integrated intensity as a function of laser power in 1L-WSe$_2$/4L-MoS$_2$ HS.

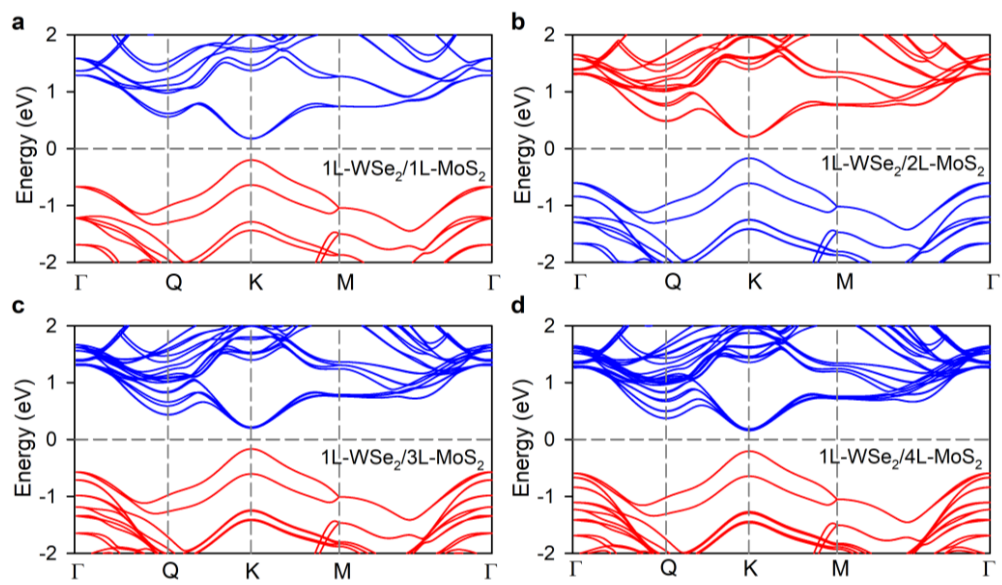

**Fig. S15 | The evolution of electronic energy structure of the 1L-WSe$_2$/mL-MoS$_2$ HS with the number of MoS$_2$ layers. a-d,** The calculated electronic energy structure of the 1L-WSe$_2$/1L-MoS$_2$ HS, the 1L-WSe$_2$/4L-MoS$_2$ HS, the 1L-WSe$_2$/3L-MoS$_2$ HS, and the 1L-WSe$_2$/4L-MoS$_2$ HS by using the Perdew-Burke-Ernzerhof-type (PBE) method, respectively. The valance bands at the K point are mainly from the WSe$_2$ layers, and the conduction bands at the K point are mainly from the MoS$_2$ layers. Both the conduction band minimum (CBM) at K and Q point decrease as MoS$_2$ layer number is increased from 1 to 4. Although the decrease at Q point is larger than in K point, the K point is still lower than Q point in 1L-WSe$_2$/mL MoS$_2$ (m=1-4) HS. Consequently, 1L-WSe$_2$/mL MoS$_2$ HS is a direct band gap semiconductor. With MoS$_2$ layer number is continuously increased, the Q point will overtake the K point to be the CBM of the HS. We should note that the calculated bandgap is underestimated.

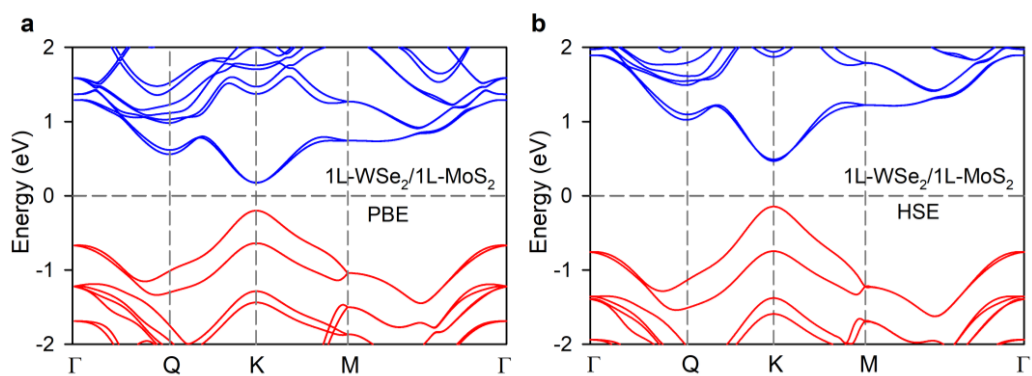

**Fig. S16 | The electronic energy structure of the 1L-WSe$_2$/1L-MoS$_2$ HS. a,** The calculated electronic energy structure of the 1L-WSe$_2$/1L MoS$_2$ HS by using the Perdew-Burke-Ernzerhof-type (PBE) method as the exchange-correlation functional. Here, the band gap is around 0.377 eV. **b,** The calculated electronic energy structure of 1L-WSe$_2$/1L MoS$_2$ HS by using a more sophisticated Heyd−Scuseria−Ernzerhof hybrid functional method (HSE) as the exchange–correlation potential. Here, the band gap is around 0.612 eV,

which is closer to the experimental value.

## Section V. Circularly polarized PL spectra of other mL-WSe$_2$/1L-MoS$_2$ and 1L-WSe$_2$/mL-MoS$_2$ HS samples

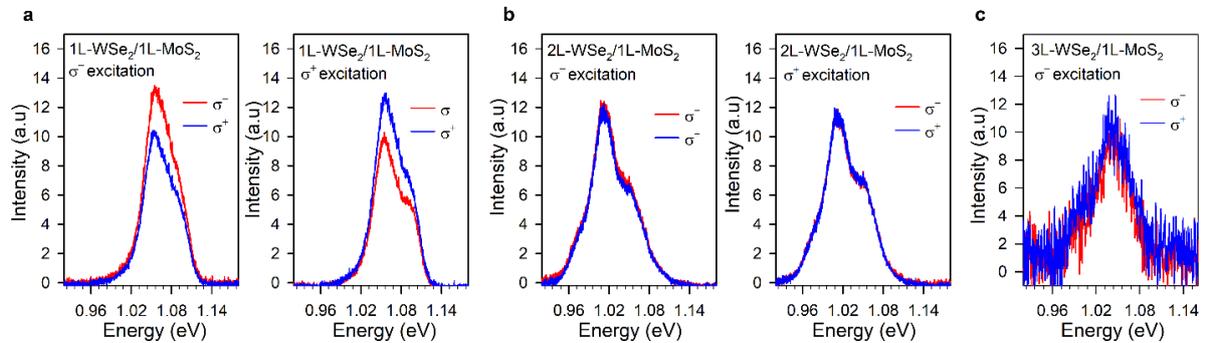

**Fig. S17 |** The circularly polarized PL spectra of IXs from the 1L-WSe$_2$/1L-MoS$_2$, 2L-WSe$_2$/1L-MoS$_2$ and 1L-WSe$_2$/3L-MoS$_2$ HS under different circular polarization configurations, respectively.

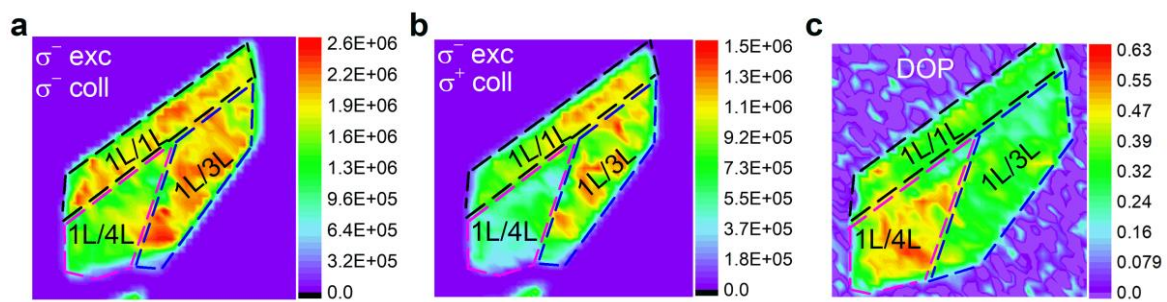

**Fig. S18 | Observation of highly polarized interlayer excitons in the 1L-WSe2/mL-MoS$_2$ HS. a,b,** The PL intensity mapping of IXs from the 1L-WSe$_2$/mL-MoS$_2$ HS (sample S2) at 4.3 K under σ$^-$ and σ$^+$ collection polarization configurations, respectively. **c,** The calculated valley DOP mapping of IXs based on (a) and (b).

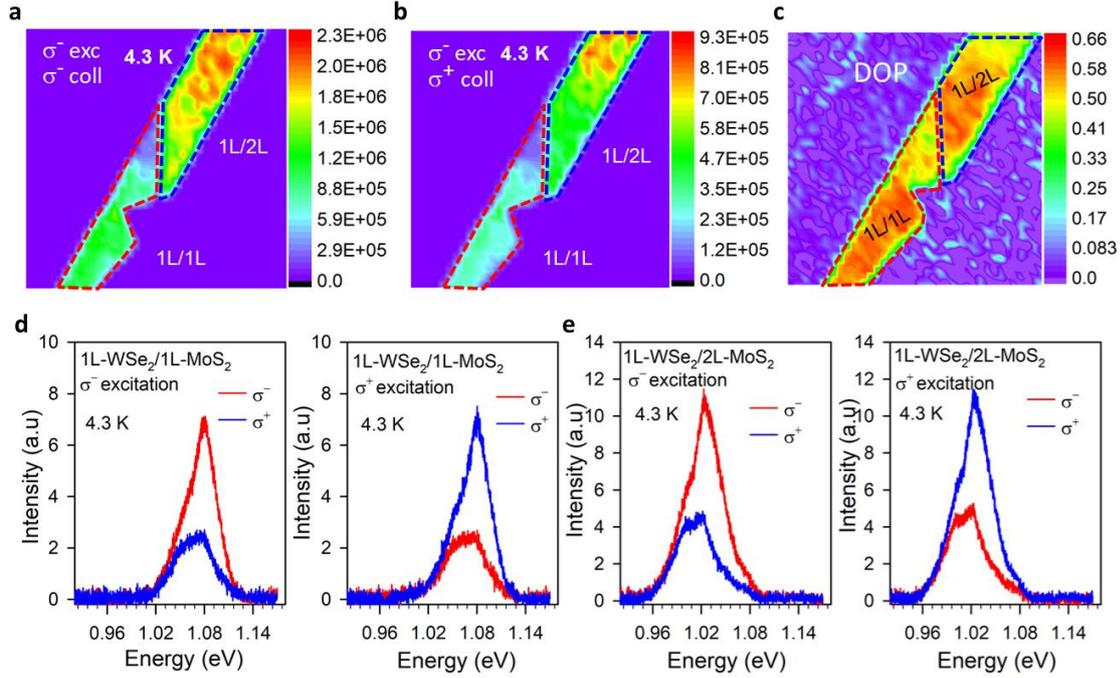

**Fig. S19 | Observation of highly polarized interlayer excitons in the 1L-WSe$_2$/mL-MoS$_2$ HS. a,b,** The PL intensity mapping of IXs from the 1L-WSe$_2$/mL-MoS$_2$ HS (sample S5) at 4.3 K under σ$^-$ and σ$^+$ collection polarization configurations, respectively. **c,** The calculated valley DOP mapping of IXs based on (a) and (b). **d-e,** The circularly polarized PL spectra of IXs from the 1L-WSe$_2$/1L-MoS$_2$ and 1L-WSe$_2$/2L-MoS$_2$ HS under different circular polarization configurations, respectively.

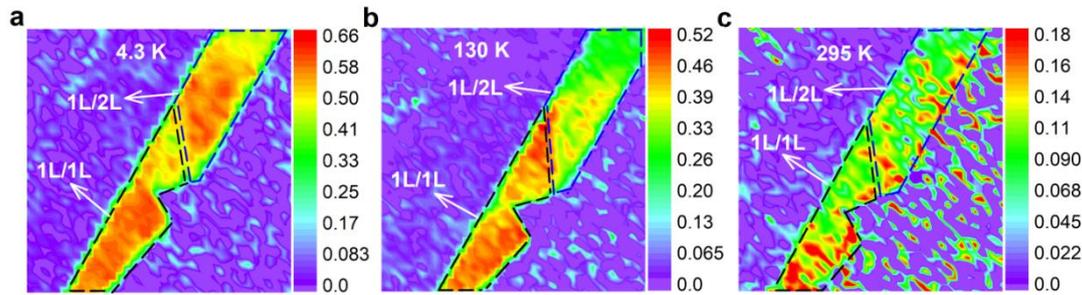

**Fig. 20 | Temperature-dependent valley polarization of interlayer exciton in the multilayer WSe$_2$/MoS$_2$ heterostructures. a-c**, The valley DOP mapping of IXs from the 1L-WSe$_2$/1L-MoS$_2$ and 1L-WSe$_2$/2L-MoS$_2$ (sample 2) at 4.3 K, 130 K, and 295 K, respectively. The inhomogeneity in (c) is from the relatively low signal-to noise ratio (SNR) at room temperature.

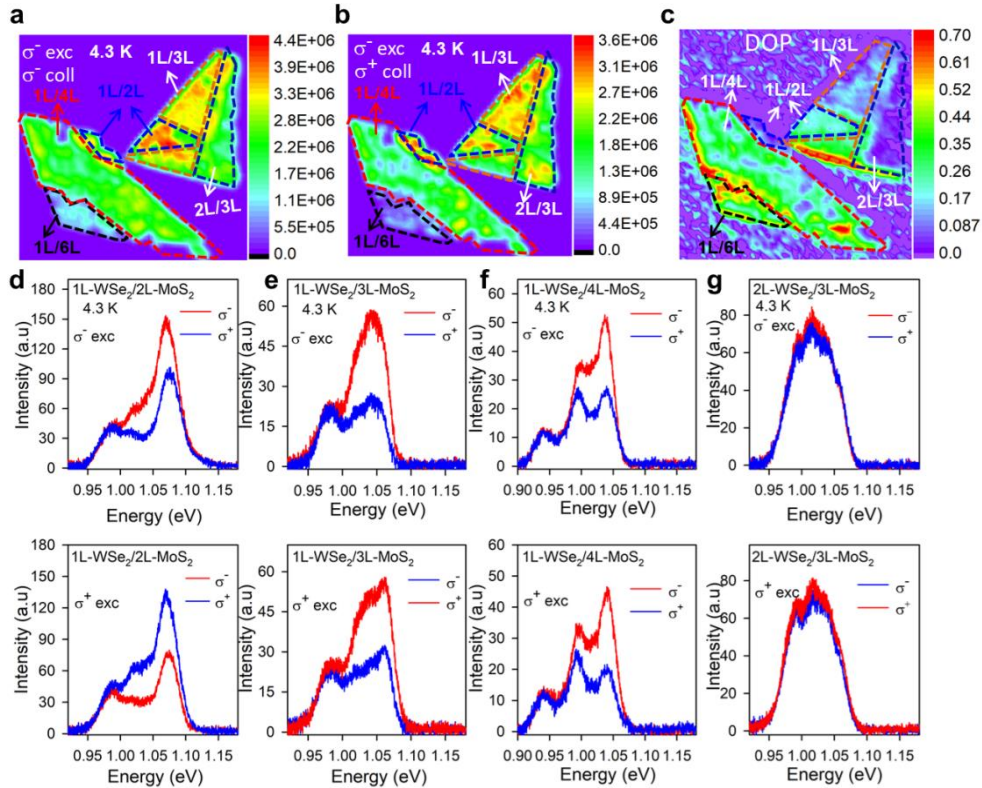

**Fig. S21 | Observation of highly polarized interlayer excitons in the 1L-WSe$_2$/mL-MoS$_2$ HS. a,b,** The PL intensity mapping of IXs from the 1L-WSe$_2$/mL-MoS$_2$ HS (sample S6) at 4.3 K under σ$^-$ and σ$^+$ collection polarization configurations, respectively. **c,** The calculated valley DOP mapping of IXs based on (a) and (b). **d-g,** The circularly polarized PL spectra of IXs from the 1L-WSe$_2$/2L-MoS$_2$, 1L-WSe$_2$/3L-MoS$_2$, 1L-WSe$_2$/4L-MoS$_2$, and 2L-WSe$_2$/3L-MoS$_2$ HS under different circular polarization configurations.

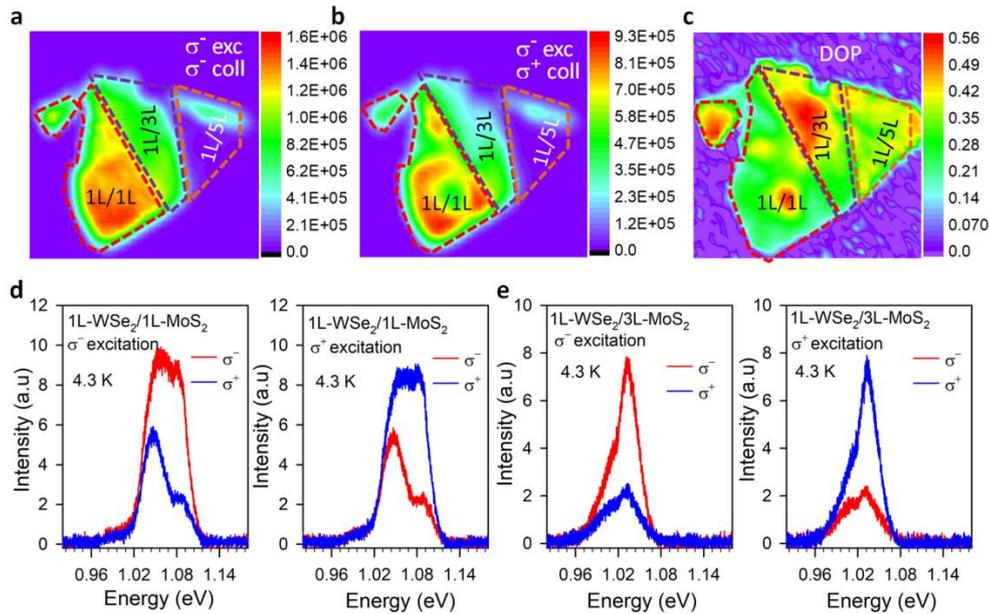

**Fig. S22 | Observation of highly polarized interlayer excitons in the 1L-WSe$_2$/mL-MoS$_2$ HS. a,b,** The PL intensity mapping of IXs from the 1L-WSe$_2$/mL-MoS$_2$ HS (sample S7) at 4.3 K under σ$^-$ and σ$^+$ collection polarization configurations, respectively. **c,** The calculated valley DOP mapping of IXs based on (a) and (b). **d-e,** The circularly polarized PL spectra of IXs from the 1L-WSe$_2$/1L-MoS$_2$ and

1L-WSe$_2$/3L-MoS$_2$ HS under different circular polarization configurations, respectively.

## Section VI. Time-resolved PL decay of IXs in mL-WSe$_2$/1L-MoS$_2$ HS and valley lifetime of IX in mL-WSe$_2$/1L-MoS$_2$.

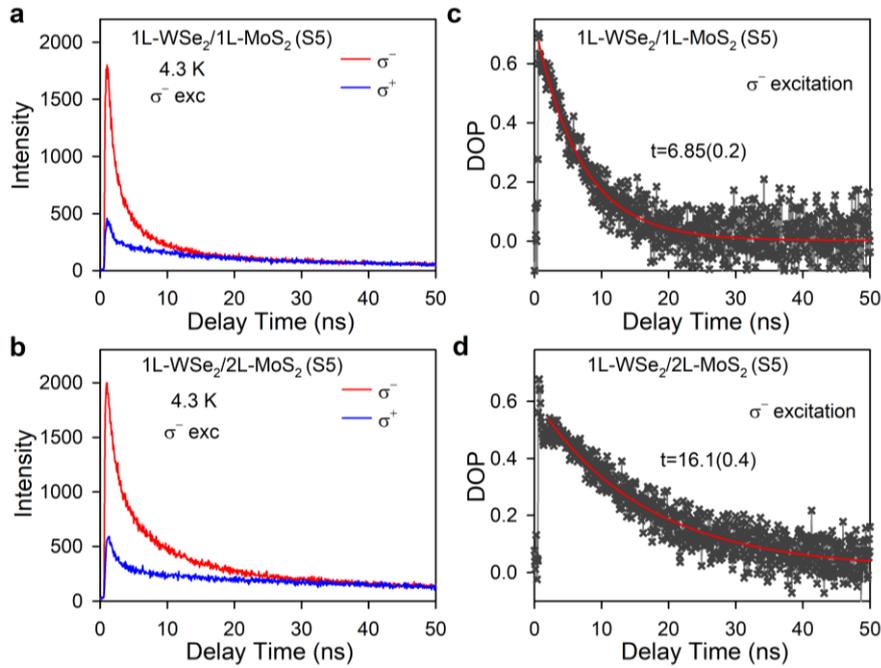

**Fig. S23 | Layer dependent DOP lifetime of the interlayer excitons in the 1L-WSe$_2$/2L-MoS$_2$ HS. a, b,** The time-resolved PL of IXs from the 1L-WSe$_2$/1L-MoS$_2$ and 1L-WSe$_2$/2L-MoS$_2$ HS (sample S5) under left circularly polarized light excitation at 4.3 K. **c, d,** The time-resolved valley DOP of IXs. The value in the parentheses represents the fitting error.

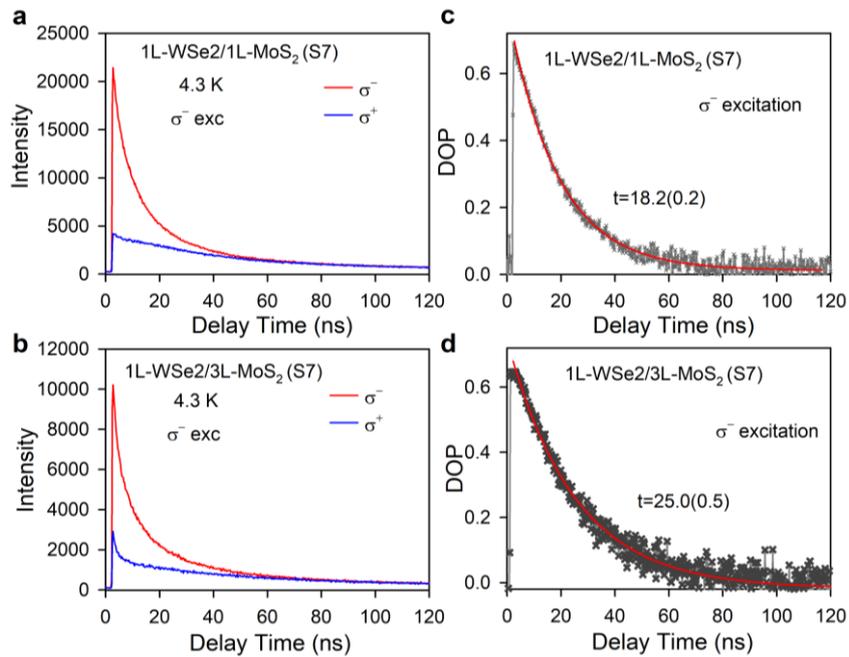

**Fig. S24 | Layer dependent DOP lifetime of the interlayer excitons in 1L-WSe$_2$/3L-MoS$_2$ HS. a, b,** The

time-resolved circularly polarized PL of IXs (Sample 7) at 4.3 K. **c, d,** The time-resolved DOP of IXs. The value in the parentheses represents the fitting error.

## Section VII. Temperature dependence of valley lifetimes

Figure S26 shows the depolarization lifetime under $\sigma^-$ and $\sigma^+$ excitation as a function of temperature. In contrast to the case in 1L/1L heterostructure, where the valley DOP lifetime only has a weak dependence with the temperature in this temperature ranges, the valley DOP lifetime in 1L/3L and 1L/4L heterostructure regions decays quickly with increasing temperature. This result can be understood by considering the role of phonon in valley depolarization. When the temperature increases, more phonons are activated to participate in the intervalley scattering process, which leads to faster hole valley depolarization. In the 1L/mL (m>1) heterostructure, due to more atoms in the unit cell and the interlayer coupling, more phonons compared with that in 1L/1L heterostructure are participating in the electron and hole intervalley scattering process. Therefore, the valley depolarization lifetime in 1L/3L and 1L/4L heterostructure regions are dropping faster than that in the 1L/1L heterostructure region at the elevated temperature.

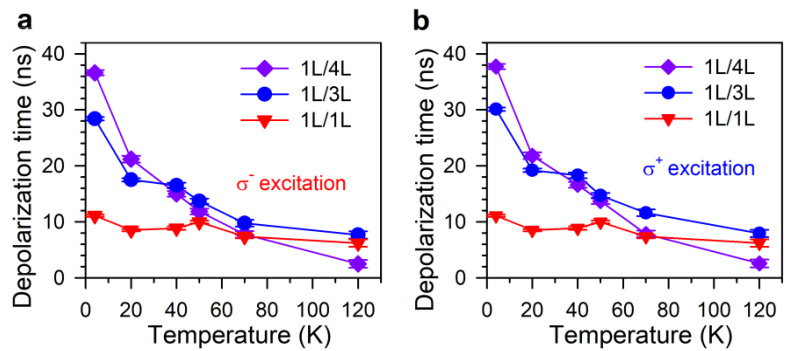

**Fig. S25 | Temperature dependence of valley lifetime of interlayer excitons in 1L-WSe$_2$/mL-MoS$_2$ HS. a, b,** The valley lifetime as a function of temperature in 1L-WSe2/mL-MoS2 (m=1, 3, 4) HS.

# Section VIII. Magnetic field dependence of circularly polarized PL spectra and valley lifetime of IXs in mL-WSe$_2$/1L-MoS$_2$.

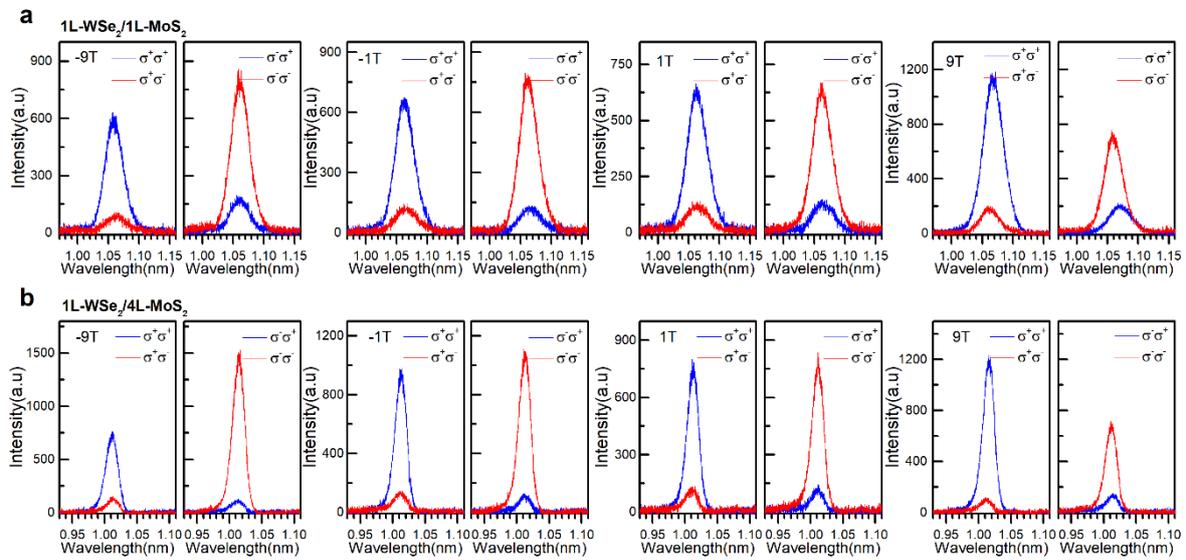

**Fig. S26 | Magnetic field dependence of the circularly polarized PL spectra of interlayer excitons in 1L-WSe$_2$/mL-MoS$_2$ HS. a,** The circularly polarized PL spectra of IXs in 1L-WSe$_2$/1L-MoS$_2$ HS under -9T, -1T, 1T and 9T, respectively. **b,** The circularly polarized PL spectra of IXs in 1L-WSe$_2$/4L-MoS$_2$ HS under -9T, -1T, 1T and 9T, respectively.

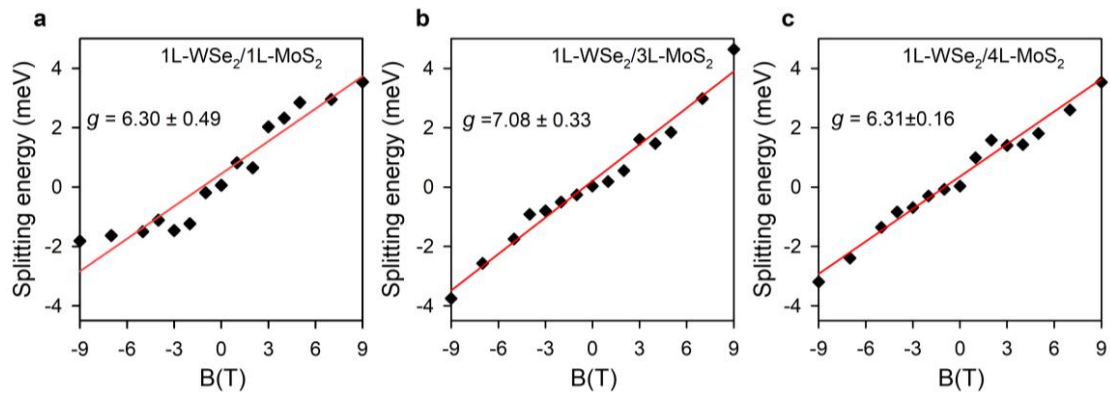

**Fig. S27 | Valley splitting of the interlayer excitons in the 1L-WSe$_2$/mL-MoS$_2$ HS. a-c,** The valley splitting of IXs as a function of magnetic fields in 1L-WSe$_2$/mL-MoS$_2$ (m=1, 3, 4) HS. The g factor is around 6-7 in all HS regions, implying the AA stacking order.

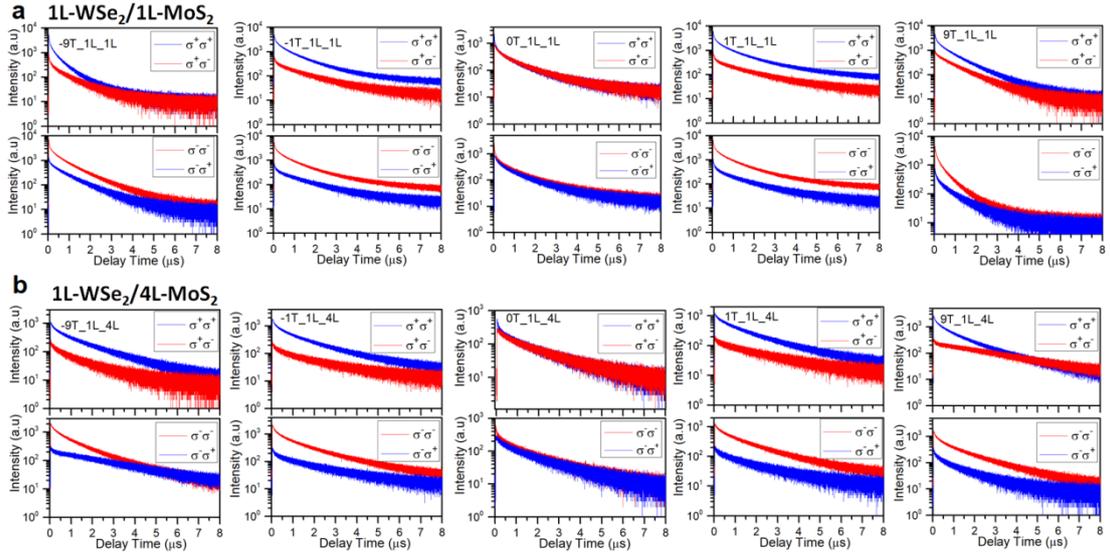

**Fig. S28 | Circularly polarized PL decay of the interlayer excitons under magnetic fields in the 1L-WSe$_2$/mL-MoS$_2$ HS. a,** 1L-WSe$_2$/1L-MoS$_2$ case under -9T, -1T, 1T and 9T, respectively. **b,** 1L-WSe$_2$/4L-MoS$_2$ case under -9T, -1T, 1T and 9T, respectively.

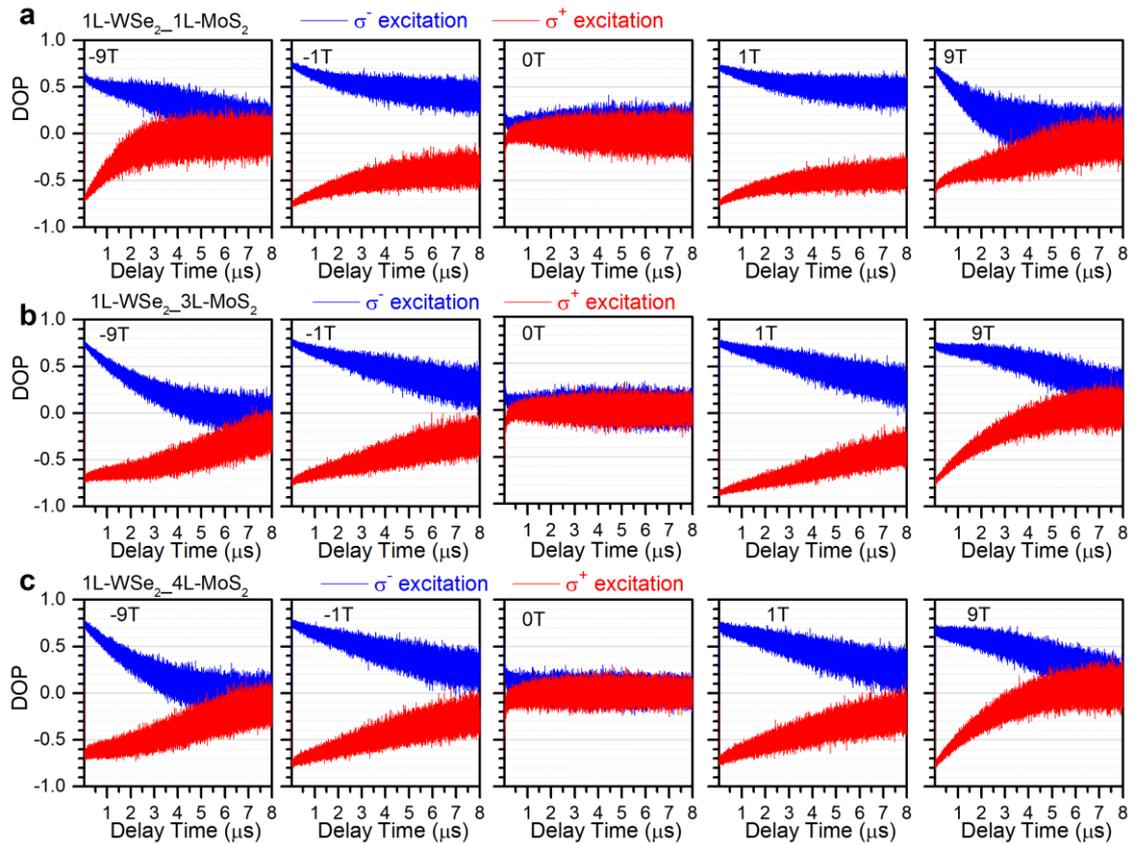

**Fig. S29 | Magnetic dependence of DOP decay of the interlayer excitons in the 1L-WSe$_2$/mL-MoS$_2$ HS. a-c,** The valley lifetime of IXs in 1L-WSe$_2$/1L-MoS$_2$, 1L-WSe$_2$/4L-MoS$_2$, 1L-WSe$_2$/4L-MoS$_2$ HS with right- and left circular polarization excitation at different magnetic fields (-9T, -1T, 0T, 1T and 9T), respectively.